\documentstyle [12pt]{article}

\newcommand{\etal}{{et al.}~}

\newcommand{\de}{\delta}
\newcommand{\te}{\theta}
\newcommand{\la}{\lambda}
\newcommand{\p}{\partial}
\newcommand{\f}{\frac}

\newcommand{\Om}{\Omega}

\newcommand{\s}{\sigma}
\newcommand{\al}{\alpha}

\newcommand{\bfx}{{\bf x}}

\newcommand{\bfv}{{\bf v}}

\newcommand{\calO}{{\cal O}}

\newcommand{\calK}{{\cal K}}
\newcommand{\calN}{{\cal N}}

\newcommand{\calP}{{\cal P}}

\newcommand{\bc}{\begin{center}}
\newcommand{\be}{\begin{equation}}
\newcommand{\ee}{\end{equation}}
\newcommand{\ec}{\end{center}}
\newcommand{\lan}{\langle}
\newcommand{\ran}{\rangle}
\newcommand{\spose}[1]{\hbox to 0pt{#1\hss}}
\newcommand{\lta}{\mathrel{\spose{\lower 3pt\hbox{$\mathchar"218$}}
 \raise 2.0pt\hbox{$\mathchar"13C$}}}
\newcommand{\gta}{\mathrel{\spose{\lower 3pt\hbox{$\mathchar"218$}}
 \raise 2.0pt\hbox{$\mathchar"13E$}}}

\input{epsf}
\begin{document}

\title{ Weakly Nonlinear Density-Velocity Relation}
\vspace{0.5cm}
\author{
{\bf Micha{\l} J. CHODOROWSKI}$^{\,1}$ \\
and \\
{\bf Ewa L. {\L}OKAS}$^{\,2}$ \\ \\
{\it Copernicus Astronomical Center} \\
{\it Bartycka 18, 00--716 Warsaw, Poland } \\
$^{1}$ e-mail: michal@camk.edu.pl\\
$^{2}$ e-mail: lokas@camk.edu.pl\\}
%
\maketitle
\section*{Abstract}

We rigorously derive weakly nonlinear relation between cosmic density
and velocity fields up to third order in perturbation theory.  The
density field is described by the mass density contrast, $\de$.  The
velocity field is described by the variable $\te$ proportional to the
velocity divergence, $\te = - f(\Omega)^{-1} H_0^{-1}
\nabla\cdot\bfv$, where $f(\Omega) \simeq \Omega^{0.6}$, $\Omega$ is
the cosmological density parameter and $H_0$ is the Hubble constant.
Our calculations show that mean $\de$ given $\te$ is a third order
polynomial in $\te$, $\lan \de \ran|_{\te} = a_1 \te + a_2 (\te^2 -
\s_\te^2) + a_3 \te^3$. This result constitutes an extension of the
formula $\lan \de \ran|_{\te} = \te + a_2 (\te^2 - \s_\te^2)$, found
by Bernardeau~(1992) which involved second order perturbative
solutions. Third order perturbative corrections introduce the cubic
term. They also, however, cause the coefficient $a_1$ to depart from
unity, in contrast with the linear theory prediction. We compute the
values of the coefficients $a_p$ for scale-free power spectra, as well
as for standard CDM, for Gaussian smoothing. The coefficients obey a
hierarchy $a_3 \ll a_2 \ll a_1$, meaning that the perturbative series
converges very fast. Their dependence on $\Omega$ is expected to be
very weak. The values of the coefficients for CDM spectrum are in
qualitative agreement with the results of N-body simulations by Ganon
\etal (1996). The results provide a method for breaking the
$\Omega$-bias degeneracy in comparisons of cosmic density and velocity
fields such as IRAS-POTENT.

\vspace{0.5cm}
\noindent{\bf Key Words:} cosmology: theory -- galaxies: clustering --
galaxies:  formation -- large--scale structure of the Universe

\section{Introduction}

The most common assumption in theory of structure formation is the
gravitational instability hypothesis: the observed large-scale
structure has formed by the gravitational amplification of
small-amplitude fluctuations present in the primordial density
field. Cosmic velocity fields of galaxies result consequently from the
gravitational attraction of large-scale mass inhomogeneities, that
perturb the uniform Hubble flow. The quantitative relation between the
peculiar velocity field, $\bfv$, and the mass density contrast field,
$\de = \rho / \rho_b -1$, where $\rho_b$ is the background density,
can be inferred from the dynamical equations for the pressureless
self-gravitating cosmic fluid.

In linear regime, i.e. for \ $\de \ll 1$, the fluctuation field grows
in time by an overall scale factor $D(t)$ (which depends on the
cosmological parameter $\Omega$), preserving its initial shape, $
\de(\bfx,t) = D(t)\,\de(\bfx,t_i)$. As a result, the linear theory
relation between the density and the velocity field is {\em local} \be
\de(\bfx) = - f(\Omega)^{-1} H_0^{-1} \nabla \cdot \bfv(\bfx) \,,
\label{e1}
\ee
where $H_0$ is the Hubble constant and $f(\Omega) \simeq \Omega^{0.6}$
(see e.g.\ Peebles 1980). One can use the above formula to reconstruct
from the large-scale velocity field the (linear) mass density field,
up to an $\Omega$-dependent multiplicative factor $f(\Omega)$. The
comparison of the reconstructed mass field with the observed
large-scale {\em galaxy} density field could therefore serve as a test
for the gravitational instability hypothesis and as a method for
estimating $\Omega$ (Dekel \etal 1993).

There are, however, both observational evidence and theoretical
arguments for thinking that galaxies are biased tracers of mass.
When the fluctuations are small one can assume that the galaxy and
mass density contrast fields are linearly related, $\de_g = b\,\de $,
hence
\be
- \f{1}{H_0} \nabla \cdot \bfv(\bfx) = \f{f(\Omega)}{b} \de_g(\bfx) \,.
\label{e2}
\ee
The comparison of the POTENT-reconstructed mass field with the IRAS
galaxy field yields $f(\Omega)/b_{{\rm IRAS}}$ values close to unity
(Dekel \etal 1993, Dekel 1994). It is then tempting to conclude that
the large-scale dynamics is consistent with an assumption of $\Omega =
1$, provided that $b_{{\rm IRAS}}$ is also close to unity. However,
since we do not know anything a priori about bias, we should measure
it independently.

It has been suggested that nonlinear corrections to the linear
density-velocity relation (hereafter DVR), equation~(\ref{e1}), can
help to perform such a measurement (Yahil 1991). The corrections are
indeed necessary because there are points in the $\de_{{\rm
POTENT}}$-$\de_{{\rm IRAS}}$ correlation diagram for which the density
contrast reaches unity, clearly contradicting the underlying
assumption of $\de \ll 1$. On the other hand the rms fluctuation of
the mass field, $\sigma$, is smaller (but not much smaller) than unity
that means that the field is {\em weakly nonlinear}.

In weakly nonlinear regime perturbation theory can be efficiently
applied, as the results of N-body simulations show (Juszkiewicz \etal
1995, Bernardeau 1994a,b, Baugh, Gazta{\~n}aga \& Efstathiou 1995,
{\L}okas \etal 1995, Bernardeau \& van de Weygaert 1996). However,
most of the attempts to derive a weakly nonlinear extension of the
linear DVR have been based on the Zel'dovich approximation and its
modifications (Nusser \etal 1991, Gramman 1993). The Zel'dovich
approximation is a useful qualitative guess of nonlinear dynamics but
it provides only approximate answers for rigorously derived
higher-order perturbative solutions. Consequently, it does much better
than linear theory, but still does not predict accurately the weakly
nonlinear relation between velocity and density, as verified by N-body
simulations (Mancinelli \etal 1994, Ganon \etal 1996).

The first attempt, and so far the only one, to calculate DVR within
the framework of rigorous Eulerian perturbation theory has been taken
up by Bernardeau~(1992a; hereafter B92). B92 has calculated the exact
DVR for an unsmoothed final field in the limiting case of vanishing
variance. The assumption $\sigma \to 0$ greatly simplifies
mathematical calculations. It is not, however, well suited for the
application to the IRAS-POTENT comparison: we are then not interested
in the statistics of very rare events $(\de \gg\sigma)$ of linear
field ($\sigma \ll 1$), but in the statistics of `typical' events
($\de \sim \sigma$) of a weakly nonlinear field ($\sigma \lta
1$). N-body cosmological simulations show that the exact formula of
B92, when straightly applied to the case $\sigma \lta 1$, works worse
than the Zel'dovich approximation (Mancinelli \etal 1994, Ganon \etal
1996).

B92 has also computed the first nonlinear (i.e.\ quadratic) correction
for the DVR in the case of a smoothed final field with non-vanishing
variance.  However, neither the details of the derivation, nor the
explicit form of the coefficient of the corrective term are given in
the paper. On the other hand, a perturbation theory-inspired
approximation of density as a third-order polynomial of velocity
divergence turns out to be an excellent robust fit to N-body results
(Mancinelli \etal 1994, Ganon \etal 1996). Theoretical construction of
such a polynomial requires third order perturbative solutions and
provides therefore higher order corrections to the DVR than those
given by B92. All this inspired us to calculate the weakly nonlinear
DVR of a smoothed final field with $\sigma \lta 1$ up to third order
in perturbation theory.

The paper is organized as follows: in section~2 we derive weakly
nonlinear DVR in its general form. In section~3 we compute values of
the coefficients entering this relation for the case of scale-free
power spectra, as well as for standard CDM. Discussion and concluding
remarks are given in section~4.

\section{General derivation of the density-velocity relation}

In perturbation theory, one expands the solution for the density
contrast as a series around the background value $\de =0$,
\be
\de = \de_{1} + \de_{2} + \de_{3} + \ldots \,,
\label{e3}
\ee
and truncates it at some order. The linear theory solution mentioned
in section~1 is just the perturbation theory series truncated at the
lowest, i.e.~first order term,
\be \de_{1}(\bfx,t) = D(t) \,\de(\bfx,t_i) \,.
\label{e4}
\ee
Higher-order solutions are found iteratively: the second order
contribution $\de_{2}$ is the solution to the dynamical equations with
$\de_{1}$ as the source term for nonlinearities, and so on. Throughout
this paper, we will consider only the growing modes, as the remaining
ones are suppressed during linear evolution. In general, the $n$-th order
solution is found to be of the order of $(\de_1)^n$ (Fry 1984; Goroff
\etal 1986).  Let us define $\s$ as the square root of the variance of the
linear density field, i.e.~$\s^2 = \lan \de^2_1 \ran$, with $\lan \cdot
\ran$ meaning the ensemble averaging. We have $\de_1 \sim \s$, $\de_n \sim
\s^n$, and the series~(\ref{e3}) is a power series in a small parameter
$\s$.

We describe the velocity field by a variable proportional to the
velocity divergence,
\be \te(\bfx, t)\equiv - f(\Omega)^{-1} H_0^{-1}
\nabla\cdot\bfv(\bfx, t)
\label{e5}
\ee
(which is slightly different from the commonly used definition, e.g.\
Bernardeau~1994a). The variable $\te$ is as well expanded in a series
\be \te = \te_{1} + \te_{2} + \te_{3} + \ldots \,.
\label{e6}
\ee
The linear theory solution, equation~(\ref{e1}), therefore gives
\be \de_{1}(\bfx) = \te_{1}(\bfx) \,.
\label{e7}
\ee

Second order contributions to $\de$ and $\te$ are different. Their
explicit dependence on $\Omega$ is extremely weak and in the range of
cosmological interest, $0.1 \leq \Om \leq 3$, the solutions are
excellently approximated by the expressions which hold in the case of
the Einstein-de Sitter universe (Bouchet \etal 1992), namely (Goroff
\etal 1986)
\be \de_{2}(\bfx, t) =
\f{5}{7}\,\de_{1}^{\,2} + \p_{\alpha}\de_{1}\,\p_{\alpha}\Delta_{1} +
\f{2}{7}\p_{\al}\p_{\beta}\,\Delta_{1}\p_{\al}\p_{\beta}\Delta_{1}\;,
\label{e8}
\ee
and
\be \te_{2}(\bfx,t) = \f{3}{7}\,\de_{1}^{\,2} +
\p_{\alpha}\de_{1}\,\p_{\alpha}\Delta_{1} +
\f{4}{7}\p_{\al}\p_{\beta}\,\Delta_{1}\p_{\al}\p_{\beta}\Delta_{1}\;.
\label{e9}
\ee
Here, $\Delta_{1}(\bfx, t)$ is the linear gravitational potential,
\be
\Delta_1(\bfx) = - \int \f{{\rm d}^3 x'}{4\pi} \f{\de_{1}(\bfx')}{\vert
\bfx - \bfx' \vert} \;.
\label{e10}
\ee
We see therefore that up to second order in perturbation theory the
divergence of the velocity field, $\nabla\cdot\bfv(\bfx) = - f(\Omega)
H_0 (\te_{1} + \te_{2} + \cdots)$, depends explicitly on $\Omega$ only
via a multiplicative factor $f(\Omega)$.

Equation~(\ref{e7}) means that if we read from a {\em linear} field
the values of pairs ($\de(\bfx)$, $\te(\bfx)$), point by point, and
plot them on the $\de$-$\te$ plane, they will lie on a straight
line. The second order contributions to $\de(\bfx)$ and $\te(\bfx)$,
in addition to the local term $\sim \de_{1}^{\,2}(\bfx)$, contain
non-local terms due to the linear gravitational potential, $\sim \sum
\alpha_{\bfx' \bfx''}\de_{1}(\bfx') \de_{1}(\bfx'')$. As a result, the
weakly nonlinear DVR is no longer local. However, given $\te$, the
spread in the values of $\de$ comes only from nonlinear corrections.
Consequently, the points on the $\de$-$\te$ plane are still strongly
correlated: they are expected to form an elongated set of length $\sim
\s$ and width $\sim \s^2$ around the mean trend (B92). This has been
also observed in N-body simulations (Nusser \etal 1991, Bernardeau \&
van de Weygaert 1996). The mean trend can therefore serve as a very
useful local approximation of a true nonlocal DVR.

Full information about the density-velocity correlation is contained
in a joint probability distribution function (PDF) of weakly nonlinear
variables $\de$ and $\te$. Pioneering works on computing PDFs of
weakly nonlinear cosmological density and velocity fields have been
performed by Bernardeau. First, Bernardeau (1992b) computed a
one-point PDF of an unsmoothed weakly nonlinear density
field. Subsequently, he extended his calculations for the case of a
top-hat window function (Bernardeau 1994a), and computed as well the
PDF of a top-hat smoothed velocity divergence field (Bernardeau
1994b). A {\em joint} density-velocity PDF, however, is still known
only for the case of an unsmoothed final field, and in the limit $\s
\to 0$. B92 calculated it in the form of mean $\te$ given $\de$; this
relation is however easily invertible and the result is
\be
\de = \left(1 + \f{2}{3} \te \right)^{3/2} - 1 \,.
\label{e11}
\ee
Note that in the linear theory limit, $\te \ll 1$, the above equation
indeed reduces to equation~(\ref{e7}).

Juszkiewicz \etal (1995) by means of N-body simulations have shown
that a one-point PDF of a single variable $\de$ (or $\te$) in the
range of `typical' events $\de \sim \s$ can be very well approximated
by the so-called Edgeworth series (e.g.\ Longuet-Higgins~1963, 1964
and references therein). The Edgeworth series is constructed from
cumulants of the true distribution, defined as the connected part of
the moments, \be \kappa_n = \lan \de^n \ran_{conn} \,.
\label{e12}
\ee

The cumulants of order $n > 2$ provide an effective measure of
non-Gaussianity because for a Gaussian distribution they vanish.
Throughout this paper we will assume Gaussian initial conditions.
Consequently, all $n > 2$ cumulants of a fluctuation field are
initially zero. During nonlinear phase of evolution, however, they
acquire nonzero values. Fry (1984) showed that cumulants of cosmic
density and velocity fields in weakly nonlinear regime obey the
following scaling
\be
\kappa_n = S_n \s^{2(n-1)} + {\cal O}(\s^{2n})\,.
\label{e13}
\ee
To calculate the coefficient $S_n$, the perturbative solution of
($n-1$)th order is needed. Let us define dimensionless quantities
related to cumulants as follows
\be \lambda_n =
\f{\kappa_n}{\kappa_2^{n/2}} \,,
\label{e14}
\ee
where by definition $\kappa_2 = \lan \de^2 \ran = \s_\de^2$ is the
nonlinear variance of a density field. From equation~(\ref{e13}) one
can deduce the order of weakly nonlinear corrections to its linear
value
\be \s_\de^2 = \s^2 + {\cal O}(\s^{4})\,.
\label{e14a}
\ee
The coefficients $\lambda_n$ are thus the cumulants of a standardized
variable $\mu = \de/\s_\de$ and we will refer to them as to `standard
cumulants'. The first two nontrivial standard cumulants, $\lambda_3$
and $\lambda_4$, are called in statistics skewness and kurtosis,
respectively.

The Edgeworth series reads
\be
    p(\mu) = \frac{1}{\sqrt{2 \pi}} \ {\rm e}^{-\mu^{2}/2}
    \left[ 1 + \frac{1}{6} \lambda_3 H_{3}(\mu) \right.
    + \frac{1}{24} \lambda_4 H_{4}(\mu)
    \left. + \frac{1}{72} \lambda_3^2 H_{6}(\mu) + \cdots \right]
\label{e15}
\ee
where $H_{n}(\mu)$'s are the $n$--th order Hermite polynomials
generated by
\be
    (-1)^{n} \frac{{\rm d}^{n}}{{\rm d} \mu^{n}} {\rm e}^{-\mu^{2}/2} =
    {\rm e}^{-\mu^{2}/2} H_{n}(\mu) \,.
\label{e16}
\ee
In Table~1 we provide explicit forms of the few lowest order
polynomials. From Equations~(\ref{e13}), (\ref{e14}) and (\ref{e14a})
we have
\be \lambda_n = S_n \s_\de^{n-2} + {\cal O}(\s_\de^{n})\,,
\label{e17}
\ee
which expresses the scaling behaviour of standard cumulants of a
weakly nonlinear density field evolving from Gaussian initial
conditions. In particular, $\lambda_3 = S_3 \s_\de$ and $\lambda_4 =
S_4 \s_\de^2$, i.e.\ during weakly nonlinear evolution skewness and
kurtosis grow like the rms fluctuation of the field and the square of
it, respectively.  In cosmology, there is a long tradition to call
`skewness' and `kurtosis', respectively, the coefficients $S_3$ and
$S_4$ themselves.  We will honour it hereafter.

Equation~(\ref{e17}) ensures that the Edgeworth series is a series
expansion of an exact PDF in powers of a small parameter $\s_\de$
(Longuet-Higgins 1963). In weakly nonlinear regime we can thus
approximate the true PDF by the Edgeworth series truncated at some
order. Using equation~(\ref{e17}) the Edgeworth expansion,
equation~(\ref{e15}), can be rewritten in the explicitly perturbative,
third-order form
\be p(\mu) = \frac{1}{\sqrt{2 \pi}} \ {\rm
e}^{-\mu^{2}/2} \left[ 1 + \frac{1}{6} S_{3} \sigma_\de H_{3}(\mu)
\right.  + \frac{1}{24} S_{4} \sigma_\de^{2} H_{4}(\mu) \left. +
\frac{1}{72} S_{3}^{2} \sigma_\de^{2} H_{6}(\mu) \right] \,.
\label{e18}
\ee
The Edgeworth expansion for the variable $\te$ or $\nu = \te /\s_\te$,
where $\s_\te^2 = \lan \te^2 \ran$ is the nonlinear variance of the
velocity divergence field has the same form, except that $S_3$ and
$S_4$ are then the skewness and the kurtosis of the velocity
divergence field.

The third-order Edgeworth expansion describes accurately the shape of
a true PDF up to $\mu \sim \s_\de^{-1}$ (Juszkiewicz \etal 1995). The
failure of the approximation in the very tails reflects the fact that
it is constructed only from a few low order cumulants of the true
distribution (see also Bernardeau \& Kofman 1995). In the present
paper, however, we are not interested in the statistics of very rare
events in the $\de$-$\te$ space. Instead, we want to calculate just
the lowest conditional moment: mean $\de$ given $\te$. For this
purpose, we have to know the approximate form of the joint
distribution for the variables $\de$ and $\te$ that needs to be
accurate only for typical events, $\de \sim \s_\de$, $\te \sim
\s_\te $. The two-point generalization of the above third-order
Edgeworth series exactly satisfies this condition.

In fact, one can proceed in two ways. One can derive joint Edgeworth
expansion and then calculate conditional moments from it. One can
also, however, calculate the moments directly. Deriving the
third-order joint Edgeworth expansion is a straightforward, but
lengthy calculation, while of most interest for cosmology is just the
first moment, describing the mean trend. Therefore, in this paper we
calculate it directly, postponing the calculation of the joint
Edgeworth expansion and higher-order moments (e.g.~the variance around
the mean trend) resulting from it to the next paper. The methods of
calculating moments and the full PDF are still closely related and in
the following calculation we are inspired to some extent by
Longuet-Higgins~(1963), who derived a {\em second-order} joint
Edgeworth expansion in order to apply it to statistical theory of sea
waves.

The conditional probability for $\de$ given $\te$ is
\be
\left. p(\de) \right|_{\te} = \f{p(\de,\te)}{p(\te)}
\label{e19}
\ee
where $p(\de,\te)$ is the joint PDF for $\de$ and $\te$. The
characteristic function of $p(\de,\te)$ is
\be
\Phi(it,is) = \int\!\!\int e^{it\de + is\te} p(\de,\te) {\rm d}\de
{\rm d}\te \,. \label{e20}
\ee
Expanding the exponentials we obtain
\be
\Phi(it,is) = \sum_{m,n = 0}^{\infty} \f{\lan \de^m \te^n \ran}{m! n!}
(it)^m (is)^n \,,
\label{e21}
\ee
where $\lan \de^m \te^n \ran$ are the joint moments of $\de$ and
$\te$,
\be
\lan \de^m \te^n \ran = \int\!\!\int \de^m \te^n p(\de,\te)\,
{\rm d}\de {\rm d}\te \,.
\label{e22}
\ee
If the joint moments are known, $p(\de,\te)$ can be calculated via the
inverse Fourier transform,
\be
p(\de,\te) = \f{1}{(2\pi)^2} \int\!\!\int e^{-it\de -is\te} \Phi(it,is)
\, {\rm d}t {\rm d}s \,.  \label{e23}
\ee

Mean $\de$ given $\te$, $\lan \de \ran |_{\te}$, is by definition $\int
\de\,p(\de)|_{\te} {\rm d}\de $. From equation~(\ref{e19}) we have
\be
\lan \de \ran |_{\te} = \f{\int \de\,p(\de,\te)\, {\rm d}\de}{p(\te)} \,.
\label{e24}
\ee
Let us denote $\int \de\,p(\de,\te) {\rm d}\de$ by $\calN$. By
equation~(\ref{e23}),
\begin{eqnarray}
\calN
&=& \f{1}{(2\pi)^2} \int\!\!\int\!\!\int e^{-it\de -is\te} \de\,
\Phi(it,is)\, {\rm d}t {\rm d}s {\rm d}\de
\nonumber \\
&=& - \f{1}{(2\pi)^2} \int\!\!\int e^{-is\te} \Phi(it,is)\,
{\rm d}t {\rm d}s\,
\f{\p}{\p(it)} \int e^{-it\de} {\rm d}\de
\nonumber \\
&=& - \f{1}{2\pi} \int\!\!\int e^{-is\te} \Phi(it,is)\, \f{\p}{\p(it)}
\de_D(t)\, {\rm d}t {\rm d}s \,, \label{e25}
\end{eqnarray}
where $\de_D(t)$ denotes the Dirac delta function. Integrating by
parts we obtain
\be
\calN = \f{1}{2\pi} \int e^{-is\te} \f{\p}{\p(it)} \Phi(it,is)|_{t=0}
\, {\rm d}s\,.  \label{e26}
\ee

The characteristic function is related to the cumulant generating
function, $\calK$, by the equation
\be
\Phi(it,is) = \exp{[\calK(it,is)]} \,. \label{e27}
\ee
The cumulants, $\kappa_{mn}$, from which $\calK$ is constructed,
\be
\calK = \sum_{(m,n) \ne (0,0)}^\infty  \f{\kappa_{mn}}{m! n!}
(it)^m (is)^n \,,  \label{e28}
\ee
are given by the {\em connected} part of the joint moments
\be
\kappa_{mn} = \lan \de^m \te^n \ran_{conn} \,.
\label{e29}
\ee
Using equations~(\ref{e27}) and (\ref{e28}) we obtain
\be
\f{\p}{\p(it)} \left. \Phi(it,is)\right|_{t=0} =
\left[ \sum_{n=0}^\infty \f{\kappa_{1n}}{n!} (is)^n \right] \,
\exp\!\left[\sum_{n=1}^\infty \f{\kappa_{0n}}{n!} (is)^n \right] \,.
\label{e30}
\ee
By definition, $\kappa_{0n}$ are the ordinary cumulants of the
variable $\te$. The variables $\de$ and $\te$ have zero mean, so
$\kappa_{10} = \kappa_{01} = 0$. Equations~(\ref{e26}) and~(\ref{e30})
then give
\begin{eqnarray}
\calN
&=& \f{1}{2\pi} \int_{-\infty}^\infty {\rm d}s\, e^{-is\te} \left[
\sum_{n=1}^\infty \f{\kappa_{1n}}{n!} (is)^n \right]
\, \exp\!\left[\sum_{n=2}^\infty \f{\kappa_{0n}}{n!} (is)^n \right]
\nonumber \\
&=& \f{1}{2\pi} \int_{-\infty}^\infty {\rm d}s\, e^{-\te(is) +
\f{1}{2} \kappa_{02}
(is)^2} \left[ \sum_{n=1}^\infty \f{\kappa_{1n}}{n!} (is)^n \right]
\, \exp\!\left[\sum_{n=3}^\infty \f{\kappa_{0n}}{n!} (is)^n \right]
\,.  \label{e31}
\end{eqnarray}

Let us define a new variable $z = \kappa_{02}^{1/2} s$ and let us
recall that $\mu$ and $\nu$ are the standardized variables,
\be
\mu = \f{\de}{\s_\de} = \f{\de}{\kappa_{20}^{1/2}} \qquad
\hbox{and} \qquad
\nu = \f{\te}{\s_\te} = \f{\te}{\kappa_{02}^{1/2}} \,.
\label{e31a}
\ee
We then have
\be \calN = \f{1}{2\pi\kappa_{02}^{1/2}}
\int_{-\infty}^\infty {\rm d}z\, e^{-\f{1}{2}(z^2 + 2i\nu z)} \left[
\sum_{n=1}^\infty \f{\kappa_{1n}}{n!\kappa_{02}^{n/2}} (iz)^n \right]
\, \exp\!\left[\sum_{n=3}^\infty \f{\kappa_{0n}}{n!
\kappa_{02}^{n/2}} (iz)^n \right] \,.
\label{e32}
\ee
The standard joint cumulants are defined by
\be
\lambda_{mn} = \f{\kappa_{mn}}{\kappa_{20}^{m/2} \kappa_{02}^{n/2} }
\,, \label{e33}
\ee
hence, $\kappa_{0n} / \kappa_{02}^{n/2} = \lambda_{0n}$, $\kappa_{1n}
/ \kappa_{02}^{n/2} = \kappa_{20}^{1/2} \lambda_{1n}$ and
\be
\calN = \f{1}{2\pi} \left(\f{\kappa_{20}}{\kappa_{02}}\right)^{1/2}
\int_{-\infty}^\infty {\rm d}z\, e^{-\f{1}{2}(z^2 + 2i\nu z)}
\left[ \sum_{n=1}^\infty \f{\lambda_{1n}}{n!} (iz)^n \right]
\, \exp\!\left[\sum_{n=3}^\infty \f{\lambda_{0n}}{n!} (iz)^n \right] \,.
\label{e34}
\ee

One may ask why we have introduced the cumulant generating function:
using just the characteristic function $\Phi$, the above equation
would look formally simpler. The reason is similar to that in case of
constructing one-point PDF. From perturbation theory it follows that
standard joint cumulants, equation~(\ref{e33}), obey the following
scaling hierarchy
\be \la_{mn} = S_{mn} \s^{m+n-2} + {\cal O}(\s^{m+n}) \,.
\label{e35}
\ee
where $\s$ is the linear variance of $\de$ or, equivalently, of $\te$
(recall that at linear order $\de = \te$). The series in
equation~(\ref{e34}) are therefore power series in a small parameter
$\s$ and truncating them at some order $p$ we neglect contributions
which are $\sim \s^{p+1}$. Perturbation theory also predicts that
\be
\s_\te^2 = \lan \te^2 \ran = \kappa_{02} = \s^2 + {\cal O}(\s^{4}) \,,
\label{e35b}
\ee
so when we are interested in the leading order terms in
hierarchy~(\ref{e35}) we can use linear $\s$ instead of nonlinear
$\s_\te$ (or $\s_\de$, see eq.~[\ref{e14a}]).

In the present paper we want to calculate the weakly nonlinear
extension of the linear $\de$-$\te$ relation, up to cubic in $\te$,
${\cal O}(\s_\te^3)$ terms. In equation~(\ref{e34}), relating mean
$\mu = \de/\s_\de$ and $\nu = \te/\s_\te$, we will thus keep terms up
to the order of $\s_\te^2$. We have
\begin{eqnarray}
\calN
&=& \f{1}{2\pi} \left(\f{\kappa_{20}}{\kappa_{02}}\right)^{1/2}
\int_{-\infty}^\infty {\rm d}z\, e^{-\f{1}{2}(z^2 + 2i\nu z)}
\nonumber \\
&\times& \left[ \la_{11}(iz) + \f{\la_{12}}{2} (iz)^2 +
\f{\la_{13}}{6} (iz)^3 \right] \left[ 1 + \f{\la_{03}}{6} (iz)^3
+ \biggl\{ \f{\la_{04}}{24} (iz)^4 + \f{\la_{03}^2}{72} (iz)^6\biggr\}
\right] \,.
\label{e36}
\end{eqnarray}
In the expression above, $\la_{12} \sim \la_{03} \sim \s_\te$ and
$\la_{13} \sim \la_{04} \sim \la_{03}^2 \sim \s_\te^2$. The cumulant
$\la_{11}$ deserves a separate, more detailed treatment. Defined as
$\lan \de \te\ran / [\lan \de^2 \ran^{1/2} \lan \te^2 \ran^{1/2}]$
(see eq.[\ref{e33}]), it is the correlation coefficient between the
fields $\de$ and $\te$. Since the fields are identical at first order,
$\de_1 = \te_1$, at the lowest order $\la_{11} = 1$. From
equation~(\ref{e35}) it follows that the higher-order correction to this
value of $\la_{11}$ is $\calO(\s_\te^2)$, so in general $\la_{11} = 1 +
{\cal O}(\s_\te^2)$.  Multiplying the polynomials in equation~(\ref{e36})
we keep only the terms up to the order of $\s_\te^2$. It means also that
we replace the products $\la_{11} \la_{mn}$ with $m+n\ge 3$ by $\la_{mn}$,
since the correction is of at least cubic order in $\s_\te$. After sorting
the resulting terms of the form $(iz)^n$ we integrate them, using the
identity
\be
\f{1}{\sqrt{2\pi}} \int_{-\infty}^\infty {\rm d}z\,
e^{-\f{1}{2}(z^2 + 2i\nu z)} (iz)^n = H_n(\nu)\, e^{-\f{1}{2} \nu^2}
\label{e37}
\ee
where $H_{n}$ are the $n$-th order Hermite polynomials.
The result is
\begin{eqnarray}
\calN
&=& \f{1}{\sqrt{2\pi}} \left(\f{\kappa_{20}}{\kappa_{02}}\right)^{1/2}
e^{-\f{1}{2} \nu^2}
\nonumber \\
&\times& \left[\la_{11} H_1(\nu) + \f{\la_{12}}{2} H_2(\nu) +
\f{\la_{03}}{6} H_4(\nu) \right.
\nonumber \\
&+& \left. \; \f{\la_{13}}{6} H_3(\nu) + \left(\f{\la_{04}}{24} +
\f{\la_{12} \la_{03}}{12} \right) H_5(\nu) +
\f{\la_{03}^2}{72} H_7(\nu) \right] \,.
\label{e38}
\end{eqnarray}

To calculate mean $\de$ given $\te$ from equation~(\ref{e24}) we need
also one-point PDF of velocity divergence. As already stated, it is
given by the one-point Edgeworth series, equation~(\ref{e15}), for the
variable $\nu = \te/\kappa_{02}^{1/2}$
\be
    p(\nu) = \frac{1}{\sqrt{2 \pi}} \ {\rm e}^{-\nu^{2}/2}
    \left[ 1 + \frac{1}{6} \lambda_{03} H_{3}(\nu) \right.
    + \frac{1}{24} \lambda_{04} H_{4}(\nu)
    \left. + \frac{1}{72} \lambda_{03}^2 H_{6}(\nu) \right] \,.
\label{e39}
\ee
Recalling that $\mu = \de / \kappa_{20}^{1/2}$, and by
equation~(\ref{e24}), we have $\lan \mu \ran|_{\nu} = \kappa_{20}^{-1/2}
\lan \de \ran|_{\te} = \kappa_{20}^{-1/2} \calN / p(\te) \break =
(\kappa_{02}/\kappa_{20})^{1/2} \calN / p(\nu)$. From
equations~(\ref{e38}) and~(\ref{e39}), expanding the denominator,
multiplying, and keeping only the terms up to ${\cal O}(\s_\te^2)$ we
finally obtain
\be
\lan \mu \ran|_{\nu} = \la_{11} \nu + \calP_2(\nu) + \calP_3(\nu)\,,
\label{e40}
\ee
where
\be
\calP_2(\nu) = \f{\la_{12}}{2} H_2(\nu) + \f{\la_{03}}{6}\Bigl[
H_4(\nu) - \nu H_3(\nu) \Bigr]
\label{e41}
\ee
and
\begin{eqnarray}
\calP_3(\nu)
&=& \f{\la_{13}}{6} H_3(\nu) + \f{\la_{04}}{24} \Bigl[ H_5(\nu)
-\nu H_4(\nu) \Bigr] + \f{\la_{12} \la_{03}}{12} \Bigl[ H_5(\nu) -
H_2(\nu) H_3(\nu) \Bigr]
\nonumber \\
&+& \f{\la_{03}^2}{72} \Bigl[ H_7(\nu) - \nu H_6(\nu) -2 H_3(\nu)
H_4(\nu) + 2\nu H_3^2(\nu) \Bigr] \,.
\label{e42}
\end{eqnarray}
Note that $\calP_2(\nu)$ is ${\cal O}(\s_\te)$, and $\calP_3(\nu)$ is
${\cal O}(\s_\te^2)$.  (Remember that $\nu$ itself is standardized, so
$\nu \sim {\cal O}(\s_\te^0) = {\cal O}(1)$ and
$\s_\te$-dependence comes only from the standard cumulants.)

Equation~(\ref{e40}) expresses weakly nonlinear density-velocity
relation (DVR). In linear re\-gi\-me, $\s_\te \to 0$, the correlation
coefficient between $\de$ and $\te$ is unity, $\la_{11} =
1$. Moreover, in this limit $\calP_2$ and $\calP_3$ also approach
zero, so from equation~(\ref{e40}) we reobtain the linear theory
result, $\lan \mu \ran|_{\nu} = \nu$, or $\lan \de \ran|_{\te} =
\te$. The polynomials $\calP_2$ and $\calP_3$ are higher-order
corrections to this linear relation. As we took into account the
corrections up to third order in perturbation theory, we do not expect
terms of order higher than cubic in $\nu$ to appear in
equations~(\ref{e41})-(\ref{e42}). Indeed, in these equations all the
terms of the form $\nu^n$ with $n>3$ remarkably cancel out. To prove
this, we use the recurrent relation for the Hermite polynomials
\be
H_n(\nu) - \nu H_{n-1}(\nu) = - (n-1) H_{n-2}(\nu) \,,
\label{e43}
\ee
along with their explicit forms for $n=1,\ldots ,5$ (see Table~1). The
result is
\be
\calP_2(\nu) = \f{\la_{12} - \la_{03} }{2} \bigl(\nu^2 -1\bigr) \,,
\label{e44}
\ee
and
\be
\calP_3(\nu) = \f{\la_{13} - \la_{04}}{6} \bigl(\nu^3 - 3\nu\bigr) +
\f{(\la_{03} - \la_{12}) \la_{03}}{2} \bigl( \nu^3 - 2\nu \bigr) \,.
\label{e45}
\ee
We see that $\calP_2(\nu)$ and $\calP_3(\nu)$ contain quadratic and
cubic corrections in $\nu$, respectively.

Let us now deal with the joint cumulants in $\calP_2$. We have
$\la_{03} \s_\te^3 = \kappa_{03} = \lan \te^3 \ran = S_{3\te} \s_\te^4$, where
$S_{3\te}$ denotes the skewness of the variable $\te$. Thus we get
\be
\la_{03} = S_{3\te} \s_\te \,.
\label{e46}
\ee
The other, mixed cumulant in $\calP_2(\nu)$ is defined as $\la_{12}
\s_\te^3 = \lan (\de_1 + \de_2 + \cdots) (\te_1 + \te_2 + \cdots)^2
\ran$. Recalling that $\de_1 = \te_1$, we obtain
\be
\la_{12} = \left[\f{1}{3} S_{3\de} + \f{2}{3} S_{3\te} \right] \s_\te \,,
\label{e47}
\ee
where $S_{3\de}$ denotes the skewness of the variable $\de$. Thus
$\la_{12}$ is a linear combination of ordinary third-order cumulants
of the single variables $\de$ and $\te$. Equations~(\ref{e44}),
(\ref{e46}) and~(\ref{e47}) then yield
\be
\calP_2(\nu) = \f{\Delta S_3}{6} \s_\te \bigl(\nu^2 -1\bigr) \,,
\label{e48}
\ee
where
\be
\Delta S_3 = S_{3\de} - S_{3\te} \,.
\label{e49}
\ee

We can now calculate the lowest order weakly nonlinear extension of
the linear DVR. As we will show later, $\la_{11} = 1 + {\cal
O}(\s_\te^2)$.  Keeping the terms up to ${\cal O}(\s_\te)$ in
equation~(\ref{e40}) we thus have
\be
\lan \mu \ran|_{\nu} = \left[1+ {\cal O}(\s_\te^2)\right] \nu +
\f{\Delta S_3}{6} \s_\te \bigl(\nu^2 -1\bigr) + {\cal O}(\s_\te^2) \,,
\label{e50}
\ee
or
\be
\lan \mu \ran|_{\nu} = \nu + \f{\Delta S_3}{6} \s_\te \bigl(\nu^2 -1\bigr)
+ {\cal O}(\s_\te^2) \,.
\label{e51}
\ee
>From equation~(\ref{e35b}) it follows that $\s_\de = \s_\te + {\cal
O}(\s_\te^3)$, so $\mu = \de/\s_\de = \de/\s_\te + {\cal O}
(\s_\te^2)$, hence
\be
\lan \de \ran|_{\nu} = \s_\te \nu + \f{\Delta S_3}{6} \s_\te^2
\bigl(\nu^2 -1\bigr) + {\cal O}(\s_\te^3) \,.
\label{e52}
\ee
Since $\nu = \te/\s_\te$ we end up with
\be
\lan \de \ran|_{\te} = \te + \f{\Delta S_3}{6}
\bigl(\te^2 - \s_\te^2\bigr) + {\cal O}(\s_\te^3) \,.
\label{e53}
\ee

The above equation is the lowest, second order weakly nonlinear
extension of the linear DVR. Consequently, the coefficient of the
corrective term is composed from cumulants calculable at second order
($S_{3\de}$ and $S_{3\te}$), and the term is quadratic in $\te$. This
term is shifted down additionally by $\s_\te^2$. Note from
equation~(\ref{e24}) that $\int \lan \de \ran|_{\te} p(\te) {\rm d}
\te = \int \de\, p(\de,\te)\, {\rm d}\de {\rm d}\te = \int \de\,
p(\de)\, {\rm d}\de = \lan \de \ran$ (not conditional, but ordinary),
that is zero by definition.  The $\s_\te^2$ term, naturally emerging
from our calculations, precisely ensures this.

Apart from deriving the exact DVR in the case of an unsmoothed field
with vanishing variance, equation~(\ref{e11}), B92 calculated also the
second-order DVR including the effects of finite variance and
smoothing of a final field. Our result, equation~(\ref{e53}),
coincides exactly with equation~(17) of B92 with the coefficient $B =
\Delta S_3 / 6$.

We will deal now with the cumulants in the $\calP_3$ term. From
equation~(\ref{e46})-(\ref{e47}) we have
\be
\f{(\la_{03} - \la_{12}) \la_{03}}{2} = - \f{\Delta S_3\, S_{3\te}}{6}
\s_\te^2 \,.
\label{e54}
\ee
The joint cumulant $\la_{13}$, unlike $\la_{12}$, is not a linear
combination of ordinary cumulants of the single variables $\de$ and
$\te$. We have $(\la_{13} - \la_{04}) \s_\te^4 = \lan \de \te^3 \ran -
\lan \te^4 \ran = \lan (\de - \te) \te^3 \ran = \lan (\de_2 - \te_2 \,
+ \de_3 - \te_3 + \cdots) (\te_1 + \te_2 + \te_3 + \cdots)^3 \ran$ and
therefore
\be
\la_{13} - \la_{04} = \Sigma_4 \s_\te^2 \,,
\label{e55}
\ee
where
\be
\Sigma_4  = \f{3 \lan \te_1^2 \te_2 (\de_2 - \te_2) \ran +
\lan \de_1^3 \de_3 \ran - \lan \te_1^3 \te_3 \ran}{\s^6} \,.
\label{e56}
\ee
In the expression above $\lan \cdot \ran$ stands for the {\em
connected} part of the moments. Note that $\Sigma_4$ is {\em not}
equal to $S_{4\de} - S_{4\te}$: while the last two terms in
equation~(\ref{e56}) are indeed parts of the expressions for the
ordinary kurtosis of a single variable $\de$ or $\te$, respectively,
the first term is a truly mixed moment and constitutes a new quantity.
Using the results~(\ref{e54})-(\ref{e55}) in equation~(\ref{e45}) we
obtain
\be
\calP_3(\nu) = \f{\Sigma_4 - \Delta S_3\, S_{3\te} }{6} \, \s_\te^2 \nu^3 +
\left[ \f{\Delta S_3\, S_{3\te}}{3} - \f{ \Sigma_4}{2} \right] \s_\te^2 \nu
\,.
\label{e57}
\ee

Equation~(\ref{e40}) expresses weakly nonlinear extension of the
linear DVR up to ${\cal O}(\s_\te^2)$ corrections. The scaling of the
standard cumulants with $\s_\te$, equation~(\ref{e35}), ensures that
it was enough to calculate $\la_{mn}$ with $m+ n \ge 3$ at the lowest
order. The corrections to $\la_{11}$, however, are ${\cal
O}(\s_\te^2)$, so they cannot be neglected. Similarly, $\mu = \de /
\kappa_{20}^{1/2} = \de / \s_\de = (\s_\te / \s_\de) (\de / \s_\te) =
[1 + {\cal O}(\s_\te^2)] \de /\s_\te$, so the corrections to the
linear evolution of the variance of $\de$ and $\te$ should be taken
into account as well. Changing the variables in equation~(\ref{e40})
to $\de$ and $\te$, equation~(\ref{e31a}), we have
\be \lan \de
\ran|_{\te} = \left(\f{\kappa_{20}}{\kappa_{02}}\right)^{1/2} \la_{11}
\te + \s_\te [\calP_2(\te/\s_\te) + \calP_3(\te/\s_\te)] + {\cal
O}(\s_\te^4)\,.
\label{e58}
\ee
By definitions~(\ref{e33}) and~(\ref{e29}),
\be
\left(\f{\kappa_{20}}{\kappa_{02}}\right)^{1/2} \la_{11} =
\f{\kappa_{11}}{\kappa_{02}} = \f{\lan \de \te \ran}{\lan \te^2 \ran}
= \f{\lan \de \te \ran - \lan \te^2 \ran + \lan \te^2 \ran }{\lan
\te^2 \ran} = 1 + \f{\lan (\de - \te) \te \ran}{\lan \te^2 \ran}
\,,
\label{e59}
\ee
which after expanding $\de$ and $\te$ in perturbative series gives
\be
\left(\f{\kappa_{20}}{\kappa_{02}}\right)^{1/2} \la_{11} \te =
[1 + \Sigma_2 \s_\te^2] \,\te \,,
\label{e60}
\ee
where
\be
\Sigma_2 = \f{\lan \te_2 (\de_2 - \te_2) \ran +
\lan \de_1 \de_3 \ran - \lan \te_1 \te_3 \ran}{\s^4} \,.
\label{e61}
\ee
Again, $\Sigma_2$ is {\em not} equal to $\lan \de^2 \ran - \lan \te^2
\ran$. The last two terms in equation~(\ref{e61}) are indeed parts of the
expressions for nonlinear corrections to the linear evolution of
variance of $\de$ and $\te$, but the first term is a truly mixed moment
and constitutes a new quantity.

To obtain third-order weakly nonlinear DVR in its final form we
combine equations~(\ref{e48}), (\ref{e57}) and (\ref{e60}) with
(\ref{e58}). Note that $\calP_3$ contains also a term linear in
$\te$. We have
\be
\lan \de \ran|_{\te} = a_1 \te + a_2 (\te^2 - \s_\te^2) + a_3 \te^3
\,,
\label{e62}
\ee
where
\begin{eqnarray}
a_1
&=& 1 + \left[ \Sigma_2 + \f{\Delta S_3\, S_{3\te} }{3} -
\f{\Sigma_4}{2} \right] \s_\te^2 \,,
\label{e63} \\
a_2
&=& \f{\Delta S_3}{6} \,,
\label{e64} \\
a_3
&=& \f{\Sigma_4 - \Delta S_3\, S_{3\te} }{6} \,,
\label{e65}
\end{eqnarray}
with $\Sigma_2$ and $\Sigma_4$ given by equations~(\ref{e61})
and~(\ref{e56}), respectively. Equations~(\ref{e62})-(\ref{e65})
constitute the main result of this section. Note that we reobtain the
second-order DVR, equation~(\ref{e53}), when we neglect the ${\cal
O}(\s_\te^3)$ terms (i.e.\ $\sim \s_\te^2\te$ and $\sim \te^3$).

An important conclusion can be drawn immediately: the DVR of a weakly
nonlinear ($\s_\te \lta 1$) field is different from the linear theory
prediction, equation~(\ref{e7}), even for $|\te| \ll 1$. Namely, in
this case
\be
\lan \de \ran|_{\te} = a_1 \te - a_2 \s_\te^2 \,.
\label{e66}
\ee
Thus, the linear relation is shifted down, as already discussed.
What is perhaps even more interesting, the coefficient $a_1$ generally
departs from unity. The strength of this shift and departure depends
however on the particular values of the coefficients $a_n$. This is
the subject of the next section.

\section{Numerical calculations}

A brief outline of the perturbative solutions to the Newtonian equations
of motion needed for numerical calculations of the coefficients $a_{n}$
is given in Appendix~A.

The smoothing of the fields on scale $R$ is introduced by the convolution
of the density contrast (or velocity divergence) field and the filtering
function $W$
\begin{equation} \delta_{R}({\bf x},t) = \int {\rm d}^{3}y
  \,\delta({\bf y},t) \, W({\bf |x-y|}, R)  \; .\label{dv12}
\end{equation}
We perform our calculations for a Gaussian filter function which is
appropriate for observational analysis of cosmic velocity fields and
comparing them with the density fields (e.g. POTENT and IRAS-POTENT
comparison). The Fourier representation of the Gaussian window function is
given by
\begin{equation}
  W(kR)={\rm e}^{-k^{2} R^{2}/2}. \label{dv13}
\end{equation}

We assume a Gaussian distribution for the first order $\delta_{1}$ and
$\theta_{1}$ and define
\begin{equation}
  \sigma^{2}=\langle \delta_{1}^{2} \rangle = D^{2}(t) \int
  \frac{{\rm d}^3 k}{(2 \pi)^{3}} \, P(k) \, W^{2}(k R)
  \label{dv14}
\end{equation}
as the linear variance of the density (velocity divergence) field. We
assume that for $\sigma < 1$, the first few terms in the perturbative
expansion provide a good approximation of the exact solution. Since
$\delta_1$ and $\theta_{1}$ are assumed to be Gaussian random fields, all
their statistical properties as well as those of the higher order terms in
the perturbative series are determined by the power spectrum
$P(k)$, defined as
\begin{equation}  \label{dv15}
\langle \delta_1({\bf p}) \delta_1({\bf q}) \rangle \; =
\; (2 \pi)^{3} \delta_D\,({\bf p+q}) \, P(p).
\end{equation}

\subsection{The calculation of the coefficient {$a_{2}$}}

The values of skewness for density contrast and velocity divergence
fields, given to the lowest perturbative order respectively by
\begin{equation}    \label{dv17a}
  S_{3 \delta} = \frac{3 \langle \delta_{1}^{2} \delta_{2} \rangle}
  {\sigma^{4}}
\end{equation}
and
\begin{equation}    \label{dv17b}
  S_{3 \theta} = \frac{3 \langle \theta_{1}^{2} \theta_{2} \rangle}
  {\sigma^{4}} \,,
\end{equation}
depend on the assumed form of the power spectrum.

We begin by considering spectra with a power-law form
\begin{equation}    \label{dv16}
P(k) = C k^n, \ \ -3\leq n \leq 1,
\end{equation}
where $C$ is a normalization constant. For such fields, smoothed with
a Gaussian filter, the linear order contribution to the variance given by
equation (\ref{dv14}) is
\begin{equation}
  \sigma^{2} = C D^{2}(t) \frac{\Gamma(\frac{n+3}{2})}{(2 \pi)^{2}
  R^{n+3}}, \label{dv17}
\end{equation}
where $R$ is the smoothing scale.
The values of skewness are (\L okas et al. 1995)
\begin{equation}    \label{dv18}
  S_{3 \delta} = 3 \  _{2} F_{1} \left( \frac{n+3}{2}, \frac{n+3}{2},
  \frac{3}{2}, \frac{1}{4} \right) - \left( n+\frac{8}{7} \right) \
  _{2} F_{1} \left( \frac{n+3}{2}, \frac{n+3}{2}, \frac{5}{2}, \frac{1}{4}
  \right)
\end{equation}
\begin{equation}   \label{dv19}
  S_{3 \theta} = 3 \  _{2} F_{1} \left( \frac{n+3}{2}, \frac{n+3}{2},
  \frac{3}{2}, \frac{1}{4} \right) - \left( n+\frac{16}{7} \right) \
  _{2} F_{1} \left( \frac{n+3}{2}, \frac{n+3}{2}, \frac{5}{2}, \frac{1}{4}
  \right)
\end{equation}
where $_{2} F_{1}$ is the hypergeometric function.
Therefore the coefficient $a_{2}$ is
\begin{equation}     \label{dv20}
  a_{2} = \frac{S_{3 \delta} - S_{3 \theta}}{6} =
  \frac{4}{21} \ _{2} F_{1} \left( \frac{n+3}{2}, \frac{n+3}{2},
  \frac{5}{2}, \frac{1}{4} \right).
\end{equation}

The result is very weakly dependent on the value of the $\Omega$
parameter (for details see the appendix in \L okas et al. 1995;
Bernardeau et al. 1995 and Bouchet et al. 1992).  A good approximation
for the $\Omega$ dependence in the range $0.1 \le \Omega \le 3$ is
obtained by replacing the constant coefficient 4/21 in equation
(\ref{dv20}) with an $\Omega$-dependent function \begin{equation}
\label{dv21} G(\Omega) = \f{1 - 2 C(\Omega) + K(\Omega)}{3} \end{equation}
where
\begin{equation}    \label{dv22}
  K(\Omega) = \frac{3}{7} \Omega^{-2/63}, \ \ \
  C(\Omega) = \frac{3}{7} \Omega^{-1/21}.
\end{equation}
The second column of Table~2 gives the values of $a_{2}$ for
integer and half-integer values of the spectral index $n$ and $\Omega =
1$ while Figure~1 shows the coefficient $a_{2}$ as a function of $n$ for
three different values of $\Omega$.

We have chosen the scale-free spectra of the form (\ref{dv16})
not only because of their simplicity but also for their straightforward
applicability to realistic power spectra. Indeed, in the case of higher
order cumulants the value of the cumulant (the skewness or the kurtosis)
is very well approximated by the result for the scale-free spectra with
the effective index defined as (Bernardeau 1994a)
\begin{equation}    \label{dv23}
  n_{eff} = - \frac{R}{\sigma^2} \frac{d \sigma^{2}(R)}{d R} - 3.
\end{equation}

As an example of a scale-dependent power spectrum we consider the standard
CDM spectrum
\begin{equation}     \label{dv24}
  P(k) = \frac{C k^{n}}{\left\{ 1 + \left[ l_{1} k/\Gamma + \left(
  l_{2} k/\Gamma \right)^{3/2}+\left( l_{3} k/\Gamma \right)^{2}
  \right]^{\nu} \right\}^{2/\nu}}
\end{equation}
with $n=1$, $\Gamma = 0.5$, $\nu = 1.13$ and the constants in units
of $h^{-1}$ Mpc are $l_{1}=6.4$, $l_{2}=3.0$, $l_{3}=1.7$ (e.g.
Efstathiou, Bond \& White, 1992).  We normalize the spectrum so that the
linear rms density fluctuation in spheres of radius $R=8 h^{-1}$ Mpc is
equal to unity. Thus the definition of variance (\ref{dv14}) together with
the following shape of the spherical top hat window function in Fourier
space
\begin{equation}             \label{dv25}
  W_{TH}(kR)= 3 \sqrt{\frac{\pi}{2}} (k R)^{-3/2}
  J_{3/2}(k R)
\end{equation}
(where $J$ is the Bessel function)
and the power spectrum (\ref{dv24}) yield the normalization constant of
$C = 4.09 \times 10^{5} \ (h^{-1}$ Mpc$)^{4}$. Note, that the
normalization procedure is the only place where we use the top hat filter,
all other calculations are performed for a Gaussian window function
(\ref{dv13}).

We have calculated the coefficient $a_{2}$ for the CDM spectrum at
different smoothing scales $R$ in two ways. First we have found
numerically the exact values of $S_{3 \delta}$ and $S_{3 \theta}$ for CDM
spectrum following the procedure of \L okas et al. (1995). These values
are shown in the third and fourth column of Table~3. They produce the
exact value of the coefficient $a_{2}$ given in the fifth column of the
Table. The second option was to use the formula (\ref{dv20}) with
$n_{eff}$ corresponding to each scale, calculated from equation
(\ref{dv23}). The second column of Table~3 gives the values of the
effective index corresponding to each of the scales. Thus obtained value
of $a_{2}$ is given in the last column of Table~3. The comparison between
the two values of $a_{2}$ shows clearly that the discrepancy between them
is less than 1\% at all scales. The exact values of $a_{2}$ for the
CDM are repeated in Table~4, which summarizes the results for this
spectrum.

\subsection{The calculation of the coefficient {$a_{3}$}}

As we have shown in the previous section, the coefficient $a_{3}$ is given
by
\begin{equation}     \label{dv26}
a_{3} = \frac{\Sigma_{4} - \Delta S_{3} S_{3 \theta}}{6}
\end{equation}
where
\begin{equation}   \label{dv27}
\Sigma_{4} = \frac{3 \langle \delta_{1}^{2} \delta_{2} \theta_{2} \rangle
- 3 \langle \theta_{1}^{2} \theta_{2}^{2} \rangle + \langle \delta_{1}^{3}
\delta_{3} \rangle - \langle \theta_{1}^{3} \theta_{3}
\rangle}{\sigma^{6}}.
\end{equation}
We have named the quantity $\Sigma_{4}$ to stress its similarity to the
kurtosis of density and velocity divergence fields, which to lowest order
in perturbation theory are given respectively by
\begin{equation}        \label{dv28}
S_{4 \delta} = \frac{6 \langle \delta_{1}^{2} \delta_{2}^{2} \rangle
+ 4 \langle \delta_{1}^{3} \delta_{3} \rangle}{\sigma^{6}}
\end{equation}
and
\begin{equation}        \label{dv29}
S_{4 \theta} = \frac{6 \langle \theta_{1}^{2} \theta_{2}^{2} \rangle
+ 4 \langle \theta_{1}^{3} \theta_{3} \rangle}{\sigma^{6}}
\end{equation}
(unless the initial conditions are non-Gaussian: for details in the
latter case see Chodorowski \& Bouchet 1996). This shows that most of
the expressions constituting the value of $\Sigma_{4}$ for power law
spectra has already been calculated by \L okas et al. (1995) while
performing the calculations for kurtosis. Since they were not
published, we give them in Table~5 for integer and half integer values
of the spectral index $n$. They will also be needed in the
calculations of the next subsection. The only unknown part of
$\Sigma_{4}$ is the expression of the form $\langle \delta_{1}^{2}
\delta_{2} \theta_{2} \rangle$ which is calculated in Appendix~B.

In the case of no smoothing (when window function $W(k R) = 1$), which
also corresponds to putting the spectral index $n=-3$, a completely
analytic result can be obtained fairly easily; we get $\langle
\delta_{1}^{2} \delta_{2} \theta_{2} \rangle/\sigma^{6} = 1768/441 \approx
4.01$. In this case we have
\begin{equation} \label{dv34}
   a_{3} = - \frac{40}{3969} \approx -0.0101
\end{equation}

The sixth column of Table~5 gives the values of $\Sigma_{4}$ calculated
according to (\ref{dv27}). Finally, the third column of Table~2 lists the
values of $a_{3}$ obtained from equation (\ref{dv26}) in the case of
power-law spectra for integer and half integer values of the spectral
index $n$.

Bernardeau (1994a) showed that third order solutions for $\de$ and
$\te$, similarly to second order, depend very weakly on
$\Omega$. Since the coefficient $a_3$ is constructed from terms up to
third order, its expected dependence on $\Omega$ is very weak.
Bernardeau (1994a) did not give explicit forms for weakly
$\Omega$-dependent third order solutions. Unlike to the case of the
coefficient $a_2$ (see previous subsection), we cannot therefore
verify in detail the above supposition. Still, we are able to do this
at least for the case of the spectral index $n = -3$. The exact
formula of B92, equation~(\ref{e11}), describes the limiting case
$\s^2 \to 0$, $n = -3$ and $\Omega =0$. One cannot thus use it to
deduce the value of the coefficient $a_1$ which includes corrections
${\cal O}(\s^2)$. On the other hand, one can use this formula to
derive the values of the coefficients $a_2$ and $a_3$, which are
calculated in the limit $\s^2 \to 0$. The Taylor expansion of
equation~(\ref{e11}) yields
\begin{equation} \label{dv34a}
   a_{2}(\Omega = 0) = \frac{1}{6} \approx 0.167
\end{equation}
and
\begin{equation} \label{dv34b}
   a_{3}(\Omega = 0) = - \frac{1}{54} \approx -0.0185 \,.
\end{equation}
The corresponding values for $a_2$ and $a_3$ in the case $n = -3$,
$\Omega = 1$ are respectively $0.190$ and $-0.0101$ (Table~2). The
relative change of the value of the coefficient $a_3$ is therefore
greater than the relative change of $a_2$. Nevertheless, also $a_3$
depends on $\Omega$ extremely weakly in a sense that it almost
vanishes {\em both} for $\Omega = 1$ and $\Omega = 0$. All
kurtosis-type quantities entering the definition of $a_3$ are of order
of unity (see Table~5) and very precise cancellation of them is needed
to assure $a_3 \ll 1$. Therefore even weak dependence of the
perturbative solutions on $\Omega$ could in principle destroy this
`fine tuning'. For $n = -3$ this is clearly not the case.

In fact we were able to check it rigorously for all values of $n$ for
the kurtosis-type quantities in $\Sigma_4$ (eq.~[\ref{dv27}]) that
involve only second order solutions. In the range of $0.1 < \Omega <
3$ we found that the $\Omega$-dependence of $ \langle \delta_{1}^{2}
\delta_{2} \theta_{2} \rangle$ and $\langle \theta_{1}^{2}
\theta_{2}^{2} \rangle$ is similar and almost cancels out when the two
are subtracted. Analogously, Bernardeau (1994a) noted that the
combination $S_{4\te}/S^2_{3\te}$ is almost independent on $\Omega$,
to much bigger extent than the moments $S_{3\te}$ and $S_{4\te}$
themselves. Very weak dependence of $a_3$ on $\Omega$ is an
interesting problem and we will address it in more detail elsewhere.

To obtain the values of the coefficient $a_{3}$ for the CDM spectrum we
apply the effective index method described in the previous subsection. For
each of the indices calculated from equation (\ref{dv23}) for a given
Gaussian smoothing scale we interpolate the value of $a_{3}$ from the
values given in Table~2 using an accurate polynomial fit. The results are
presented in the last column of Table~4.

\subsection{The calculation of the coefficient {$a_{1}$}}

It has been proved that the coefficient $a_{1}$ is given by
\begin{equation}    \label{dv35}
  a_{1} = 1 + \left[ \Sigma_{2} + \frac{\Delta S_{3} S_{3 \theta}}{3} -
  \frac{\Sigma_{4}}{2} \right] \sigma_\te^{2},
\end{equation}
where
\begin{equation}    \label{dv36}
  \Sigma_{2} = \frac{ \langle \delta_{2} \theta_{2} \rangle - \langle
  \theta_{2}^{2} \rangle + \langle \delta_{1} \delta_{3} \rangle -
  \langle \theta_{1} \theta_{3} \rangle}{\sigma^{4}}
\end{equation}
and $\Sigma_{4}$ has been defined and discussed in the previous subsection.

The quantities involved in $\Sigma_{2}$ are of the form of the lowest
order weakly nonlinear corrections to the variance of the density and
velocity divergence fields which are of the order of $\sigma^{4}$
(\L okas et al. 1996)
\begin{equation}            \label{dv37}
  \f{\s^2_\de - \s^2}{\s^4} = \frac{\langle \delta_{2}^{2} \rangle +
  2 \langle \delta_{1} \delta_{3} \rangle}{\sigma^{4}}
\end{equation}
\begin{equation}            \label{dv38}
  \f{\s^2_\te - \s^2}{\s^4} = \frac{\langle \theta_{2}^{2} \rangle +
  2 \langle \theta_{1} \theta_{3} \rangle}{\sigma^{4}}.
\end{equation}
We recall that $\s^2_\de$ and $\s^2_\te$ stand for nonlinear variance
of density and velocity divergence respectively while $\sigma^{2}$ is
the linear variance given by equation (\ref{dv14}). In
equation~(\ref{dv35}) $\s_\te^2$ can be replaced by $\s^2$ since their
difference is already ${\cal O}(\s_\te^4)$.

The details of calculations of the terms involved in $\Sigma_{2}$ are
given in Appendix~C. As discussed in the Appendix, some of the terms are
divergent at spectral indices $n > -1$. Instead of dwelling on those
divergences we focused our attention on analysis concerning scale-free
power spectra in the case of no smoothing and the case of $-2 \le n < -1$
with smoothing, which is well justified observationally. As recent
analyses of measurements suggest (Gazta\~{n}aga 1994; Feldman, Kaiser \&
Peacock 1994; Peacock \& Dodds 1994) the linear power spectrum can be
approximated over large range of scales by a power law of spectral index
$n = -1.4 \pm 0.1$.

Calculated at the lowest order, the values of cumulants of an
unsmoothed field do not depend on the underlying power spectrum and
are equal to the values calculated for a smoothed field with spectral
index $n=-3$. This is not, however, the case for higher-order corrections
to their values. In the case of scale-free power spectra (\ref{dv16}) when
no smoothing is applied ($W(k R) = 1$) we obtain (see Appendix~C)
\begin{equation}             \label{dv46}
   \Sigma_{2} = \frac{1297}{4410} + h(n) \approx 0.3
\end{equation}
where $h(n)$ is the part weakly dependent on the spectral index $n$
which increases the rational number by roughly 10\%. The remaining
terms in the expression for the coefficient $a_1$,
equation~(\ref{dv35}), are skewness and kurtosis-related quantities
and it was sufficient to calculate them at the lowest order. Combining
equation~(\ref{dv46}) with the values of these terms corresponding to
the no smoothing case (i.e. the values for $n=-3$ in Table~2 and
Table~5 and the skewness values from equation (\ref{dv19})) we finally
get
\begin{equation}              \label{dv47}
   a_{1} \approx 1 - 0.4 \ \sigma^{2}.
\end{equation}

When smoothing is introduced, for $n=-2$ we obtain (see Appendix~C)
\begin{equation}      \label{dv48}
   \Sigma_{2} = \frac{23}{196} \pi \approx 0.369.
\end{equation}
Combined with the other numbers calculated for $n=-2$ this result yields
\begin{equation}              \label{dv49}
   a_{1} = 1 - 0.172 \ \sigma^{2}.
\end{equation}

In the range $-2 \le n < -1$ we calculate $\Sigma_{2}$ numerically and
$\Sigma_{4}$ by interpolating the values given in Table~5. The calculation
provides an independent check of the result for $n=-2$, equation
(\ref{dv48}), obtained analytically.  Table~6 shows the two corrections to
$a_{1}$ separately:  while the part containing $\Sigma_{4}$ remains
roughly constant, $\Sigma_{2}$ grows with $n$ until it blows up at $n=-1$.
The last column of Table~6 lists the values of $a_{1}$ in the
$\sigma$-dependent way in order not to obscure the results by choosing
arbitrary normalization needed for estimating $\sigma$.  Since the value
of $\sigma^{2}$ is of the order of unity on the scales of interest, it is
clear from Table~5 that at the observationally preferred spectral index $n
\approx -1.4$ the value of $a_{1}$ significantly departs from unity. It
must be noted, however, that the nonlinear correction strongly depends on
$n$ and reaches zero between $n=-1.6$ and $n=-1.7$.

To provide an example of the values of $a_{1}$ we have normalized the
power law spectra so that linear rms fluctuation in spheres of radius
$8\, h^{-1}$ Mpc is equal to unity. The resulting values of
$\sigma^{2}$ and $a_{1}$ for spectral indices $n= -1.4 \pm 0.1$ at two
different Gaussian smoothing scales $R = 5 \ h^{-1}$ Mpc and $R = 12 \
h^{-1}$ Mpc are listed in Table~7.

Due to the reasons mentioned in the previous subsection we cannot
explicitly examine the dependence of the coefficient $a_1$ on
$\Omega$. Still, it is constructed from moments involving second and
third order solutions which have been proved to depend on $\Omega$
very weakly. Consequently, the expected dependence of $a_1$ on
$\Omega$ is weak.

As an example of a scale-dependent power spectrum we again adopt the
standard CDM model which, because of its behaviour at large wave-numbers
($P(k) \propto k^{-3}$), does not introduce any challenges in the
integration.  The values of $\Sigma_{2}$ can be calculated numerically
for a given smoothing scale. By combining with skewness values from
Table~3 and the interpolated kurtosis-type values from Table~5 we end up
with the coefficient $a_{1}$ for the CDM spectrum.  The values for
different smoothing scales are given in the last column of Table~4.
Although the $\Sigma_{2}$ values grow with scale (the remaining input to
the correction to $a_{1}$ remains roughly constant for this range of
scales, see the last column of Table~5), the $\sigma^{2}$ values decrease
much faster and, as we would expect for the perturbative results, at
larger (i.e. more linear) scales the coefficient $a_{1}$ approaches its
linear value, unity.

Tables 2, 4, 6 and 7 and Figures 1, 2 and 3 summarize the main results
of this section. Table~2 provides the values of the coefficients
$a_{2}$ and $a_{3}$ for power law spectra in the whole range of the
spectral index: $-3 \le n \le 1$. Those results are plotted in
Figure~1 which also shows the $\Omega$-dependence of the coefficient
$a_{2}$. The values of the coefficient $a_{1}$ for power law spectra
and the range of spectral index $-2 \le n < -1$ are given in
Table~6. The correction to unity divided by $\sigma^{2}$ is plotted in
Figure~2. Table~7 lists the numerical values of $a_{1}$ at $n=-1.4 \pm
0.1$ for two different smoothing scales when the $\sigma_{8}=1$ (top hat)
normalization is adopted. The coefficients $a_{1}$, $a_{2}$ and
$a_{3}$ for the standard CDM spectrum for a wide range of smoothing
scales are provided in Table~4.  Figure~3 shows their dependence on
smoothing radius in the weakly nonlinear range of scales.

\section{Disentangling $\bf \Omega$ and linear bias}

The bottom line of our calculations is to propose a method, based on
nonlinear corrections to the linear DVR, for measuring independently
$\Omega$ and bias from an IRAS-POTENT-like, density-velocity
comparison. Let us assume that the galaxy and mass density contrast
fields are linearly related, i.e.\ $\de_g = b\,\de$, or
\be
\de = b^{-1} \de_g \,.
\label{o3}
\ee
We introduce a new variable
\be
\de_v = - \f{1}{H_0} \nabla\cdot\bfv \,.
\label{o1}
\ee
By definition~(\ref{e5}) we have
\be
\te = f^{-1}(\Omega)\, \de_v \,.
\label{o2}
\ee

For the sake of simplicity let us consider the case in which a cosmic
field is smoothed over a sufficiently large volume that the third
order corrections to the weakly nonlinear DVR, equation~(\ref{e62}),
can be neglected. Using equations~(\ref{o3}) and~(\ref{o2}) we can
rewrite DVR in the form relating two observables: the galaxy
density contrast, $\de_g$, and the (minus) divergence of the velocity
field, $\de_v$. We have
\be
\lan \de_g \ran_{|\de_v} = \f{b}{f} \de_v + a_2 \f{b}{f^2}
\bigl(\de_v^2 - \s_v^2\bigr)
\,.
\label{o4}
\ee

In the previous section we showed that the coefficient $a_2$
practically does not depend on $\Omega$. One can thus propose
the following method for disentangling the effects of $\Omega$ and
linear bias. First, as the output of POTENT take simply $\de_v$ (i.e.\
without any corrections for nonlinearity). Next, plot the diagram
$\de_g$--$\de_v$. Finally, fit to the points a second order
polynomial,
\be
\de_g = c_1 \de_v + c_2 (\de_v^2 - \s_v^2) \,.
\label{o6}
\ee
Comparing equation~(\ref{o6}) with~(\ref{o4}) we see that the fitted
coefficients $c_1$ and $c_2$ are related to $f$ and $b$ by
\be
c_1 = \f{b}{f} \,,
\label{o7b}
\ee
and
\be
c_2 = a_2 \f{b}{f^2} \,.
\label{o8b}
\ee
So far, only the linear coefficient $c_1$ has been measured. The
results are usually expressed in terms of the variable (Dekel \etal
1993)
\be
\beta =c_1^{-1} \,.
\label{o8a}
\ee

The difficulty that from the linear density-velocity comparison one
can estimate only the ratio $f(\Omega)/b$, or $b/f(\Omega)$, is
sometimes called the `$\Omega$-bias degeneracy problem'. In our
opinion, however, there is nothing `degenerated' in the fact that one
cannot infer the values of two variables from only one equation
involving them. True degeneration would happen if in the formula for
the coefficient $c_2$ the parameters $b$ and $f$ entered only as
a ratio, e.g.\ as $(b/f)^2$. Clearly, it is not the case. We can
therefore solve equations~(\ref{o7b})-(\ref{o8b}) separately for $b$
and $f$. The result for $f$ is
\be
f(\Omega) = a_2 \f{c_1}{c_2} = a_2 \beta^{-1} c_2^{-1} \,.
\label{o10}
\ee

Unfortunately, the assumption of purely linear bias can be seriously
questioned. Indeed, the value of the IRAS skewness is merely a half of
the predicted one if $b = 1$ and this discrepancy is commonly
attributed not to linear bias but to the fact that the IRAS survey
systematically underestimates the density of galaxies in the cores of
rich clusters (Bouchet \etal 1993). It simply means that the IRAS
galaxies are nonlinearly (anti)biased tracers of mass distribution. In
general, there is no a priori reason for the assumption of linear
bias, justified for small (linear) fluctuations, to hold also in the
case of fluctuations that are weakly nonlinear. Therefore, a correct
method for estimating $\Omega$ from the comparison of the weakly
nonlinear galaxy density with velocity fields should take into account
the effects of nonlinear bias. Such a method has been invented by
Bernardeau (private communication) and will be presented in the
follow-up paper (Bernardeau, Chodorowski \& {\L}okas, in preparation).

\section{Summary and concluding remarks}

In the present paper we derived a weakly nonlinear relation between
cosmic density and velocity fields. In linear theory, the mass density
contrast, $\de$, and the velocity divergence, $\nabla\cdot\bfv$, are
in a given point linearly related. If the fields are nonlinear the
density-velocity relation (DVR) is neither linear nor local. In weakly
nonlinear regime, however, the spread around the mean trend is so
small that the conditional mean can serve as a very useful local
approximation of a true nonlinear DVR.

We computed mean $\de$ given $\te:= - f(\Omega)^{-1} H_0^{-1}
\nabla\cdot\bfv$, that is $\lan \de \ran|_{\te}$, up to third order in
perturbation theory.  According to our calculations, it is given by a
third order polynomial in $\te$, $\lan \de \ran|_{\te} = a_1 \te + a_2
(\te^2 - \s_\te^2) + a_3 \te^3$.  This formula constitutes therefore an
extension of the formula $\lan \de \ran|_{\te} = \te + a_2 (\te^2 -
\s_\te^2)$, found by Bernardeau~(1992), which involved second order
perturbative solutions. Third order perturbative corrections not only
introduce the cubic term but cause the $a_1$ coefficient to depart
from unity as well, in contrast with the linear theory prediction. We
computed the numerical values of the coefficients $a_p$ for power-law
spectra, as well as for standard CDM.  The coefficients obey the
hierarchy $a_3 \ll a_2 \ll a_1$, which means that the perturbative
series converges very fast.

The key point of the method for disentangling $\Omega$ and bias lies in
the fact that the coefficients $a_p$ are practically $\Omega$-independent.
We have shown that the dependence of the coefficient $a_{2}$ on $\Omega$
is extremely weak. We have also given some arguments for the assumption
that this is also the case for the coefficients $a_{1}$ and $a_{3}$. The
detailed analysis of this problem will be given elsewhere.

Recently Ganon \etal (1996) have performed a set of N-body simulations for
a CDM family of models in order to test different local approximations to
DVR in weakly nonlinear regime. They tested, among others, an
approximation of $\de$ as a third order polynomial in $\te$. By visual
inspection of the plots they provide one can see that the approximation
works excellently for $|\de|$ less than unity. It does not surprise us,
since it is just the regime of applicability of perturbation theory. The
values of the fitted parameters are in qualitative agreement with our
perturbative calculations for a standard CDM: $a_1$ is slightly greater
than unity, $a_2 \simeq 0.3$ and $a_3$ is equal to a few hundredths.

In order to derive accurately the values of $a_p$ from N-body one
should, however, treat properly the final velocity field, determined
in a simulation only at a set of discrete points (final positions). A
two-step smoothing procedure, commonly used, leads to rather
substantial discrepancies between N-body simulations and analytical
perturbative calculations of higher-order reduced moments ({\L}okas
\etal 1995). Recently, Bernardeau \& van de Weygaert (1996) proposed a
new method for accurate velocity statistics estimation, based on the
use of the Voronoi and Delaunay tessellations (adapted for a top-hat
window function, however). The method proved to recover the tails of
the velocity divergence distribution very accurately.  Since the
coefficients $a_p$ are given by skewness and kurtosis-like quantities
(see section~2), probing the tails of the density and velocity
distribution, the application of the method is necessary to recover
the accurate values of the coefficients from simulations. This will be
the subject of the follow-up paper (Bernardeau, Chodorowski \&
{\L}okas, in preparation).

\section*{Acknowledgments}

We wish to thank Francis Bernardeau and Adi Nusser for valuable comments
on the draft version of this paper. MC acknowledges hospitality of Alain
Omont and Fran{\c c}ois Bouchet at Institut d'Astrophysique de Paris.
EL{\L} acknowledges support of the Foundation for Polish Science. This
research has been supported in part by the Polish State Committee for
Scientific Research grants No.  2P30401607 and 2P03D00310, the French
Ministry of Research and Technology within the program PICS/CNRS No.~198
(Astronomie Pologne) and Human Capital and Mobility program of the
European Communities (ANTARES).

\section*{References}

\begin{description}

\item Baugh, C. M., Gazta{\~n}aga, E. \& Efstathiou, G. 1995, MNRAS,
  274, 1049
\item Bernardeau, F. 1992a,
   ApJ, 390, L61 (B92)
\item Bernardeau, F. 1992b,
   ApJ, 392, 1
\item Bernardeau, F. 1994a,
   ApJ, 433, 1
\item Bernardeau, F. 1994b,
   A\&A, 291, 697
\item Bernardeau, F., Juszkiewicz, R., Dekel, A. \& Bouchet, F. R. 1995,
   MNRAS, 274, 20
\item Bernardeau, F. \& Kofman, L. 1995,
   ApJ, 443, 479
\item Bernardeau, F. \& van de Weygaert, R. 1996,
   MNRAS, 279, 693
\item Bouchet, F. R., Juszkiewicz, R., Colombi, S. \& Pellat, R. 1992,
   ApJ, 394, L5
\item  Chodorowski, M. J. \& Bouchet, F. R. 1996,
   MNRAS, 279, 557

\item Dekel, A. 1994, Annu. Rev. Astron. Astrophys., 32, 371
\item Dekel, A., Bertschinger, E., Yahil, A., Strauss, M., Davis, M. \&
   Huchra, J. 1993, ApJ, 412, 1
\item Efstathiou, G. P., Bond, J. R. \& White, S. D. M. 1992, MNRAS, 258,
   1P
\item Feldman, H. A., Kaiser, N. \& Peacock, J. A. 1994,
   ApJ, 426, 23
\item Fry, J. N. 1984,
   ApJ, 279, 499
\item Ganon, G., Dekel, A., Mancinelli, P. J. \& Yahil, A. 1996,
   in preparation
\item Gazta\~{n}aga, E. 1994, MNRAS, 268, 913
\item Goroff, M. H., Grinstein, B., Rey, S.-J. \& Wise, M. B. 1986,
   ApJ, 311, 6
\item Gramman, M. 1993, ApJ, 405, L47
\item Juszkiewicz, R., Weinberg, D. H., Amsterdamski, P., Chodorowski, M.
   \& Bouchet, F. R. 1995,
   ApJ, 442, 39
\item \L okas, E. L., Juszkiewicz, R., Weinberg, D. H. \& Bouchet, F. R.
   1995,
   MNRAS, 274, 730
\item \L okas, E. L., Juszkiewicz, R., Bouchet, F. R. \& Hivon, E. 1996,
   ApJ, in press
\item Longuet-Higgins, M. S. 1963,
   Journal of Fluid Mechanics, 17, 459
\item Longuet-Higgins, M. S. 1964,
   Radio Science Journal of Research, 68D, 1049
\item Makino, N., Sasaki, M. \& Suto, Y. 1992,
   Phys. Rev. D, 46, 585
\item Mancinelli, P. J., Yahil, A., Ganon, G. \& Dekel, A. 1994,
   in: Proceedings of the 9th IAP Astrophysics Meeting ``Cosmic velocity
   fields", ed. F. R. Bouchet and M. Lachi\`{e}ze-Rey, (Gif-sur-Yvette:
   Editions Fronti\`{e}res), 215
\item Nusser, A., Dekel, A., Bertschinger, E. \& Blumenthal,
G. R. 1991, ApJ, 379, 6
\item Peacock, J. A. \& Dodds, S. J. 1994,
   MNRAS, 267, 1020
\item Peebles, P. J. E. 1980, The Large-scale Structure of the Universe
   (Princeton: Princeton University Press)
\item Yahil, A. 1991, in: Proceedings of the 11th Moriond Astrophysics
   Meeting ``The Early Observable Universe from Diffuse Backgrounds", ed.
   B.  Rocca-Volmerange, J.  M.  Deharveng and J. Tran Thanh Van,
   (Gif-sur-Yvette: Editions Fronti\`{e}res), 359

\end{description}

\newpage

\section*{Appendix A}

All of the perturbative calculations in Section~3 are much simpler if they
are performed in Fourier space. For the first order of the density
contrast field we have
\begin{equation} \label{dv1}
\delta_{1} ({\bf k}, t)= D(t) \int {\rm d}^{3} x \ \delta_{1}
({\bf x}) {\rm e}^{i {\bf k \cdot x}}
\end{equation}
and the inverse Fourier transform is
\begin{equation} \label{dv2}
\delta_{1} ({\bf x},t)= D(t)(2 \pi)^{-3}
\int {\rm d}^{3} k \ \delta_{1} ({\bf k})
{\rm e}^{-i {\bf k \cdot x}}.
\end{equation}

For the calculations of the coefficients $a_{n}$ the second and third
order solutions for the density contrast and velocity divergence are
needed and we give them here in the Fourier representation (e.g. Goroff
et al. 1986).  For the density field we have
\begin{equation}
  \delta_{2}({\bf k},t) = \frac{D^{2}}{(2 \pi)^3}
  \int {\rm d}^{3} p \int {\rm d}^{3} q \
  \delta_D({\bf p}+{\bf q}-{\bf k})
  \delta_{1}({\bf p}) \delta_{1}({\bf q})
  P_{2 \delta}^{(s)}({\bf p},{\bf q})  \label{dv3}
\end{equation}
\begin{equation}
  \delta_{3}({\bf k},t) = \frac{D^{3}}{(2 \pi)^{6}}
  \int {\rm d}^{3} p \int {\rm d}^{3} q \int {\rm d}^{3} r \
  \delta_D({\bf p}+{\bf q}+{\bf r}-{\bf k})
  \delta_{1}({\bf p}) \delta_{1}
  ({\bf q}) \delta_{1}({\bf r})
  P_{3 \delta}^{(s)}({\bf p},{\bf q},{\bf r}). \label{dv4}
\end{equation}
The symmetrized kernels for the density field are of the form
\begin{equation}
  P_{2 \delta}^{(s)}({\bf p},{\bf q}) = \frac{1}{14}
  J({\bf p}+{\bf q},{\bf p},{\bf q})
  \label{dv5}
\end{equation}
\begin{eqnarray}
  P_{3 \delta}^{(s)}({\bf p},{\bf q},{\bf r})=
  & A_{\delta} & \left[ \right.
  H({\bf p}+{\bf q}+{\bf r}, {\bf p})
            J({\bf q}+{\bf r},{\bf q},{\bf r}) + \nonumber \\
  & +& H({\bf p}+{\bf q}+{\bf r}, {\bf q}+{\bf r})
  \left.    L({\bf q}+{\bf r},{\bf q},{\bf r}) \right] + \nonumber \\
  &+ B_{\delta} & F({\bf p}+{\bf q}+{\bf r},{\bf p},{\bf q}+{\bf r})
            L({\bf q}+{\bf r},{\bf q},{\bf r}) + \label{dv6} \\
  & +& \left(
  \begin{array}{c}
  {\bf p} \rightarrow {\bf q} \\ {\bf q} \rightarrow {\bf r} \\
  {\bf r} \rightarrow {\bf p}
  \end{array}  \right) +
  \left(
  \begin{array}{c}
  {\bf p} \rightarrow {\bf r} \\
  {\bf q} \rightarrow {\bf p} \\
  {\bf r} \rightarrow {\bf q} \nonumber
  \end{array} \right)
  \end{eqnarray}
where $A_{\delta}=1/108$ and $B_{\delta}=1/189$.
In the expression above the notation follows that of Makino et al. (1992)
i.e.
\begin{eqnarray}
  H({\bf p},{\bf q})&=&\frac{{\bf p} \cdot {\bf q}}{q^{2}} \label{dv7} \\
  F({\bf p+q},{\bf p},{\bf q})&=& \frac{1}{2} \frac{|{\bf p+q}|^{2}
  {\bf p \cdot q}}{p^{2} q^{2}} \label{dv8} \\
  J({\bf p+q,p,q})
  &=& 4 \frac{{\bf (p \cdot q)^{2}}}{p^{2}q^{2}} +
  7 \frac{p^{2} + q^{2}}{p^{2} q^{2}}
  {\bf p \cdot q} +10 \label{dv9}\\
  L({\bf p+q,p,q})
  &=& 8 \frac{{\bf (p \cdot q)^{2}}}{p^{2}q^{2}} +
  7 \frac{p^{2}+q^{2}}{p^{2} q^{2}} {\bf p \cdot q} +6.  \label{dv10}
\end{eqnarray}

The solutions for the second and third order of the velocity divergence,
$\theta_{2}$ and $\theta_{3}$, are of the same form as the density
solutions except for the kernels $P_{2 \delta}^{(s)}$ and $P_{3
\delta}^{(s)}$ that must be replaced by the corresponding kernels $P_{2
\theta}^{(s)}$ and $P_{3 \theta}^{(s)}$. The second order kernel for the
velocity divergence is
\begin{equation}      \label{dv11}
  P_{2 \theta}^{(s)}({\bf p},{\bf q}) = \frac{1}{14}
  L({\bf p}+{\bf q},{\bf p},{\bf q}).
\end{equation}
The third order kernel for the velocity divergence is obtained from the
density kernel (\ref{dv6}) by replacing the constants $A_{\delta}$ and
$B_{\delta}$ with $A_{\theta} = 1/252$ and $B_{\theta} = 1/63$
respectively.

\section*{Appendix B}

The expression $\langle \delta_{1}^{2} \delta_{2} \theta_{2} \rangle$
can be calculated in the same way as similar expressions $ \langle
\theta_{1}^{2} \theta_{2}^{2} \rangle$ or $\langle \delta_{1}^{2}
\delta_{2}^{2} \rangle$. This is in fact an easier part of the kurtosis
calculations as it involves only the second order perturbative solutions.
We have
\begin{eqnarray}
  \frac{\langle \delta_{1}^{2} \delta_{2}
  \theta_{2} \rangle}{\sigma^{6}} & = & \frac{1}{196 \pi^{3}
  \Gamma^{3}(\frac{n+3}{2})} \int {\rm d}^{3} p \int {\rm d}^{3} q \int
  {\rm d}^{3} r \ P(p) P(q) P(r) \times \nonumber\\
  & \times & W(|{\bf p+q}|) W(|{\bf r-q}|) W(p) W(r) \times \label{dv30}\\
  & \times & J({\bf p+q,p,q}) L({\bf r-q,r,-q}) \nonumber
\end{eqnarray}
which after performing the integrations over angular variables becomes
\begin{eqnarray}
  \frac{\langle  \delta_{1}^{2} \delta_{2} \theta_{2} \rangle}{\sigma^{6}}
  & = & \frac{8
  \pi}{\Gamma^{3}(\frac{n+3}{2})} \int {\rm d} p \int {\rm d} q \int {\rm
  d} r \ (p r)^{n+3/2} q^{n+1} {\rm e}^{- p^{2} -q^{2} -r^{2}} \times
  \nonumber\\
  & \times & \left[ \frac{34}{21} I_{\frac{1}{2}}(p q) \right.
  - \left( \frac{p}{q}+\frac{q}{p} \right) I_{\frac{3}{2}}(p q)
  \left. + \frac{8}{21} I_{\frac{5}{2}}(p q) \right] \times \label{dv31}\\
  & \times & \left[ \frac{26}{21} I_{\frac{1}{2}}(q r) \right.
  - \left( \frac{q}{r}+\frac{r}{q} \right) I_{\frac{3}{2}}(q r)
  \left. + \frac{16}{21} I_{\frac{5}{2}}(q r) \right]. \nonumber
\end{eqnarray}
Expanding the Bessel functions in powers
\begin{equation}
   I_{\nu}(z) = \sum_{m=0}^{\infty} \frac{1}{m! \Gamma(\nu + m+1)}
   \left( \frac{z}{2} \right)^{\nu+2 m}  \label{dv32}
\end{equation}
and using the fact that
\begin{equation}     \label{dv33}
  \int_{0}^{\infty} p^{x} {\rm e}^{-p^{2}} {\rm d} p = \frac{1}{2}
  \Gamma \left( \frac{1+x}{2} \right)
\end{equation}
we obtain the result as a series of Gamma functions which can be summed
numerically up to arbitrary accuracy. The numerical results are given in
the third column of Table~5.

\section*{Appendix C}

Using the second order solution for the density field (\ref{dv3}) and
corresponding one for the velocity divergence we obtain
\begin{equation}     \label{dv39}
    \frac{ \langle \alpha_{2} \beta_{2} \rangle }{\sigma^{4}} =
    \frac{D^{4}(t)}{98 (2 \pi)^{6} \sigma^{4}} \int {\rm d}^{3} p \int
    {\rm d}^{3} q \ P(p) P(q) W^{2}(|{\bf p+q}| R) \ M ({\bf p+q,p,q})
\end{equation}
where $\alpha$ and $\beta$ stand for $\delta$ or $\theta$. For different
combinations we have
\begin{equation}           \label{dv39a}
  M = \left\{
\begin{array}{ll}
       J^{2} &\ \ {\rm for} \ \ \alpha = \beta = \delta \\
       J \ L &\ \ {\rm for} \ \ \alpha = \delta, \ \beta = \theta \\
       L^{2}& \ \ {\rm for} \ \ \alpha = \beta = \theta
\end{array} \right.
\end{equation}
with $J$ given by equation (\ref{dv9}) and $L$ by equation (\ref{dv10}).

The second type of terms involves the third order solution
\begin{eqnarray}
  \frac{ \langle \alpha_{1} \alpha_{3} \rangle }{\sigma^{4}}
  & = & \frac{6 D^{4}(t)}{(2 \pi)^{6} \sigma^{4}}
         \int {\rm d}^{3} p \int {\rm d}^{3} q \
         P(p) P(q) W^{2}(q R) \nonumber \\
  & \times & \left\{ \ A \ \left[  \right. \right.
         H({\bf q, - p}) \ J({\bf p+q, p, q}) \nonumber \\
  & & \left. \ \ \ \ \ \: + H({\bf q}, {\bf p+q}) \
         L({\bf p+q},{\bf p},{\bf q}) \right]        \label{dv40} \\
  & & - \ B \left. F({\bf q,-p,p+q}) \ L({\bf p+q,p,q}) \right\}.
  \nonumber
\end{eqnarray}
If $\alpha = \delta$ the constants
$A_{\delta}$ and $B_{\delta}$ must be used while if $\alpha = \theta$ they
should be replaced respectively with $A_{\theta}$ and $B_{\theta}$. The
numerical values of the constants were given after equations (\ref{dv6})
and (\ref{dv11}) respectively.

After integration the expressions can be rewritten in a general way
\begin{equation}                         \label{dv41}
  \frac{ \langle \alpha_{i} \beta_{j} \rangle }{\sigma^{4}}
  = \frac{D^4(t)}{2\pi^2\sigma^4}\int_0^{\infty} {\rm d} k \
  k^2 \ W^{2}(kR) P_{ij}(k)
\end{equation}
where
\begin{equation}           \label{dv42}
  P_{22}(k) = \frac{k^3}{98(2\pi)^2} \int_0^{\infty} {\rm d} x  P(kx)
  \int_{-1}^{+1} {\rm d} \mu P\left(k\sqrt{1 + x^2 - 2 x \mu} \right)
  \frac{ f(x, \mu)}{(1 + x^2 - 2 x \mu)^2}
\end{equation}
with
\begin{equation}           \label{dv43}
  f(x, \mu) = \left\{
\begin{array}{ll}
       (3 x + 7 \mu - 10 x \mu^{2})^{2} &\ \
       {\rm for} \ \ \alpha = \beta = \delta \\
       (3 x + 7 \mu - 10 x \mu^{2})(7 \mu - x - 6 x \mu^{2}) &\ \
       {\rm for} \ \ \alpha = \delta, \ \beta = \theta \\
       (7 \mu - x - 6 x \mu^{2})^{2} & \ \
       {\rm for} \ \ \alpha = \beta = \theta
\end{array} \right.
\end{equation}
and
\begin{equation}           \label{dv44}
   P_{13}(k) = \frac{k^3 P(k)}{(2\pi)^2}
   \int_0^{\infty}  {\rm d} x P(kx) g(x)
\end{equation}
with
\begin{equation}           \label{dv45}
  g(x) = \left\{
\begin{array}{ll}
   \frac{1}{504}\left[\frac{12}{x^2} - 158 + 100 x^2 - 42 x^4 +
   \frac{3}{x^3} (x^2 - 1)^3 (7 x^2 + 2)\ln \frac{1 + x}{|1-x|} \right] &
   \ \ {\rm for} \ \ \alpha = \beta = \delta \\
   \frac{1}{168} \left[\frac{12}{x^2} - 82 + 4 x^2 - 6 x^4 +
   \frac{3}{x^3} (x^2 - 1)^3 (x^2 + 2)\ln \frac{1 + x}{|1-x|} \right] &
   \ \ {\rm for} \ \ \alpha = \beta = \theta.
\end{array} \right.
\end{equation}

In the case of power law spectra, all the expressions (\ref{dv41}) diverge 
individually in the limit of $k \rightarrow 0$ and $kx \equiv q 
\rightarrow 0$ if $n \leq -1$.  Fortunately, as it is clear from the 
definition (\ref{dv36}), in calculating $\Sigma_{2}$ similar expressions 
are subtracted and all such diverging terms cancel out. In the opposite 
limit of $k \rightarrow \infty$, $q \rightarrow \infty$, divergencies 
occur for the terms involving second order if $n > \frac{1}{2}$ and for 
the terms containing third order if $n > -1$. As thoroughly discussed by 
\L okas et al. (1996) the only way to get along with those divergencies at 
$n > -1$ is to introduce a cutoff in the initial power spectrum at large 
wave-numbers.  The results then depend on the cutoff wave-number $k_{c}$.

For integer values of the spectral index the integrals (\ref{dv41}) can 
be performed by first finding the closed form expressions for $P_{ij}$ 
similar to those proposed by Makino et al. (1992). In the limit of large 
cutoff wave-number $k_{c} \rightarrow \infty$ a simple analytical result 
can be found by identifying the terms that dominate $P_{ij}$ in this limit 
and integrating term by term. Then the result for $n=-2$ does not depend 
on the cutoff.

\newpage

\section*{}

{\samepage
\begin{table}
\centering
\begin{tabular}{cl}  \hline \hline
  $\ell$ & \ \ \ \ \ \ \ $H_{\ell}(\nu)$\\
  \hline
0 &  1 \\
1 &  $\nu$ \\
2 &  $\nu^{2} - 1$\\
3 &  $\nu^{3} - 3 \nu$ \\
4 &  $\nu^{4} - 6 \nu^{2} + 3$ \\
5 &  $\nu^{5} - 10 \nu^{3} + 15 \nu$ \\
6 &  $\nu^{6} - 15 \nu^{4} + 45 \nu^{2} - 15$ \\
\hline
\hline
\end{tabular}
\caption{The Hermite polynomials}
\end{table}

\begin{table}
\centering
\begin{tabular}{cll}  \hline \hline
  spectral index $n$ &\ \ \ $a_{2}$  &\ \ \ $a_{3}$\\
  \hline
-3.0  &  $\frac{4}{21} \approx 0.190$ & $- \frac{40}{3969}
\approx -0.0101$\\
-2.5  &  0.192 & -0.00935  \\
-2.0  &  0.196 & -0.00548  \\
-1.5  &  0.203 & -0.000127 \\
-1.0  &  0.213 &\ 0.00713 \\
-0.5  &  0.227 &\ 0.0165    \\
\ 0\ \ & 0.246 &\ 0.0279    \\
\ 0.5 &  0.270 &\ 0.0408    \\
\ 1.0 &  0.301 &\ 0.0532    \\
\hline
\hline
\end{tabular}
\caption{
The coefficients $a_{2}$ and $a_{3}$ as functions of the spectral index $n$
for scale-free power spectra and Gaussian smoothing}
\end{table}
}
{
\samepage
\begin{table}
\centering
\begin{tabular}{rlllll}  \hline \hline
 $R$  & \ $n_{eff}$ & \ $S_{3 \delta}$ & \ $S_{3 \theta}$ & \ \ $a_{2}$ &
 $a_{2}(n_{eff})$ \\
 \hline
5   & -0.946  & 3.459 & 2.177 & 0.2136 & 0.2141  \\
10  & -0.523  & 3.305 & 1.953 & 0.2253 & 0.2262  \\
15  & -0.255  & 3.227 & 1.821 & 0.2343 & 0.2355  \\
20  & -0.0654 & 3.179 & 1.729 & 0.2416 & 0.2430  \\
50  &\ 0.462  & 3.082 & 1.482 & 0.2666 & 0.2680  \\
100 &\ 0.722  & 3.051 & 1.359 & 0.2819 & 0.2830  \\
\hline \hline
\end{tabular}
\caption{The comparison of the values of the coefficient $a_{2}$ for the
CDM spectrum calculated using the exact (fifth column) and the approximate
(sixth column) method}

\end{table}

\begin{table}
\centering
\begin{tabular}{rllll}  \hline \hline
 $R$ & \ \ \ $\sigma^{2}$ & \ \ $a_{1}$ & \ \ $a_{2} $ & \ \ $a_{3} $ \\
  \hline
5   & 0.578     & 1.196   & 0.214 & 0.00804 \\
10  & 0.121     & 1.119   & 0.225 & 0.0160 \\
15  & 0.0419    & 1.0755  & 0.234 & 0.0218  \\
20  & 0.0185    & 1.0519  & 0.242 & 0.0263  \\
50  & 0.000975  & 1.0125  & 0.267 & 0.0398  \\
100 & 0.0000803 & 1.00357 & 0.282 & 0.0465  \\
\hline \hline
\end{tabular}
\caption{The coefficients $a_{1}$, $a_{2}$ and $a_{3}$ for the CDM
spectrum. In addition, the second column provides the values of the linear
variance of the density (velocity divergence) field smoothed with a
Gaussian filter for the CDM spectrum normalized as described in the text.}
\end{table}
}
{
\samepage
\begin{table}
\centering
\begin{tabular}{clllll}  \hline \hline
  spectral index $n$ & $\langle \theta_{1}^{2} \theta_{2}^{2}
  \rangle/\sigma^{6}$ & $\langle \delta_{1}^{2} \delta_{2}
  \theta_{2} \rangle/\sigma^{6}$ & $\langle \delta_{1}^{3}
  \delta_{3} \rangle/\sigma^{6}$ & $\langle \theta_{1}^{3}
  \theta_{3} \rangle /\sigma^{6}$ & \ \ $\Sigma_{4}$ \\
  \hline
-3.0  & $\frac{1352}{441} \approx 3.07$
& $\frac{1768}{441} \approx 4.01$  & $\frac{682}{189} \approx 3.61$
& $\frac{142}{63} \approx 2.25$ & $ \frac{5536}{1323} \approx 4.18$ \\
-2.5  & 2.39   & 3.22  & 2.73  &\ 1.53   & 3.68 \\
-2.0  & 1.87   & 2.61  & 2.11  &\ 1.01   & 3.31 \\
-1.5  & 1.47   & 2.13  & 1.68  &\ 0.631  & 3.04 \\
-1.0  & 1.16   & 1.76  & 1.38  &\ 0.332  & 2.84 \\
-0.5  & 0.929  & 1.47  & 1.17  &\ 0.0799 & 2.71 \\
\ 0\ \ &0.755  & 1.24  & 1.02  & -0.155  & 2.63 \\
\ 0.5 & 0.638  & 1.06  & 0.919 & -0.398  & 2.58 \\
\ 1.0 & 0.584  & 0.916 & 0.855 & -0.677  & 2.53 \\
\hline
\hline
\end{tabular}
\caption{
The kurtosis-type quantities needed for the calculation of the
coefficients $a_{1}$ and $a_{3}$ as functions of the spectral index $n$
for scale-free power spectra and Gaussian smoothing}
\end{table}

\begin{table}
\centering
\begin{tabular}{clcl}  \hline \hline
  spectral index $n$ & \ \ $\Sigma_{2}$  & $\Delta
  S_{3} S_{3 \theta}/3 - \Sigma_{4}/2$ &\ \ \ \ $a_{1}$\\
  \hline
-2.0  & 0.369 & -0.541 & $1 - 0.172 \ \sigma^{2}$ \\
-1.9  & 0.399 & -0.532 & $1 - 0.134 \ \sigma^{2}$ \\
-1.8  & 0.440 & -0.525 & $1 - 0.0850 \ \sigma^{2}$ \\
-1.7  & 0.496 & -0.518 & $1 - 0.0217 \ \sigma^{2}$ \\
-1.6  & 0.576 & -0.511 & $1 + 0.0643 \ \sigma^{2}$ \\
-1.5  & 0.693 & -0.506 & $1 + 0.187 \ \sigma^{2}$ \\
-1.4  & 0.877 & -0.501 & $1 + 0.376 \ \sigma^{2}$ \\
-1.3  & 1.19  & -0.497 & $1 + 0.698 \ \sigma^{2}$ \\
-1.2  & 1.85  & -0.493 & $1 + 1.36 \ \sigma^{2}$ \\
-1.1  & 3.86  & -0.490 & $1 + 3.37 \ \sigma^{2}$ \\
\hline
\hline
\end{tabular}
\caption{
The contributions to the weakly nonlinear correction to the coefficient
$a_{1}$ in the most interesting range of spectral indices for power law
spectra and Gaussian smoothing}
\end{table}

\begin{table}
\centering
\begin{tabular}{crcc}  \hline \hline
  spectral index $n$ & $R$ &  $\sigma^{2} $ & $a_{1}$\\
  \hline
-1.5  &  5 & 0.656 & 1.12 \\
-1.4  &  5 & 0.638 & 1.24 \\
-1.3  &  5 & 0.620 & 1.43 \\
-1.5  & 12 & 0.176 & 1.03 \\
-1.4  & 12 & 0.157 & 1.06 \\
-1.3  & 12 & 0.140 & 1.10 \\
\hline
\hline
\end{tabular}
\caption{
The values of the coefficient $a_{1}$ for the observationally preferred
range of spectral indices of power law spectra and two different
smoothing scales}

\end{table}
}

\newpage

\begin{figure}[p]
\begin{center}
    \leavevmode
    \epsfxsize=3in
    \epsfbox[155 77 455 765]{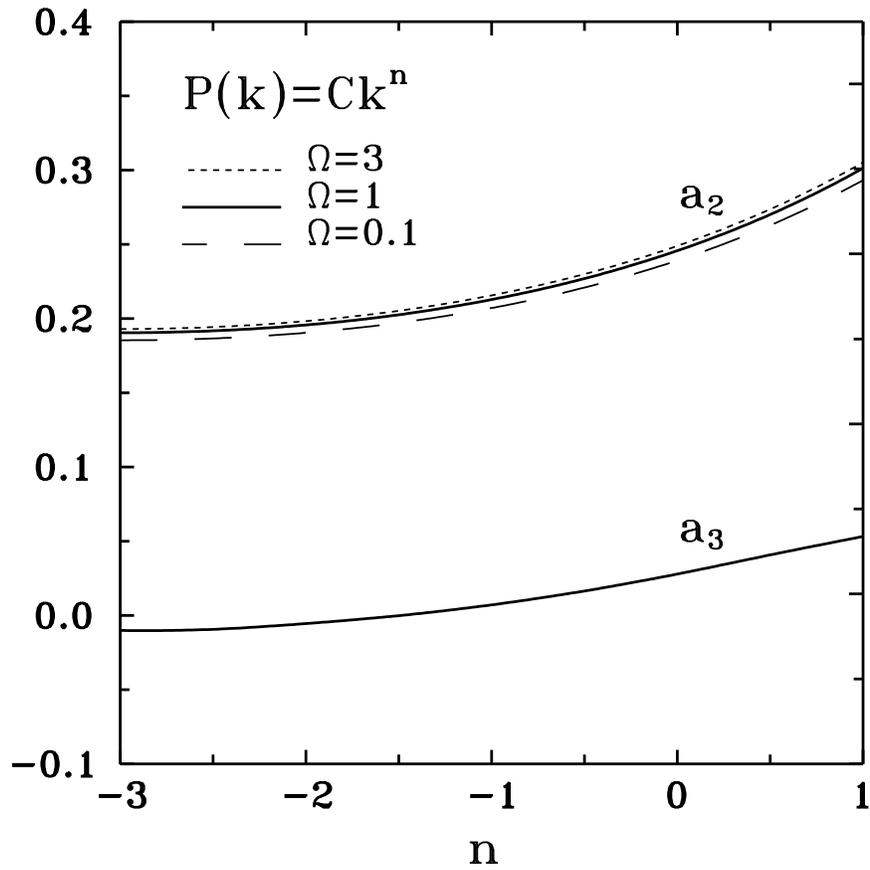}
\end{center}
    \caption{The coefficients $a_{2}$ and $a_{3}$ for scale-free power
    spectra and Gaussian smoothing as functions of the spectral index $n$.
    The solid lines correspond to the case of $\Omega = 1$. The
    coefficient $a_{2}$ is also shown for two other values of $\Omega$
    parameter (dashed lines).}
\label{dv}
\end{figure}

\begin{figure}[p]
\begin{center}
    \leavevmode
    \epsfxsize=3in
    \epsfbox[155 77 455 765]{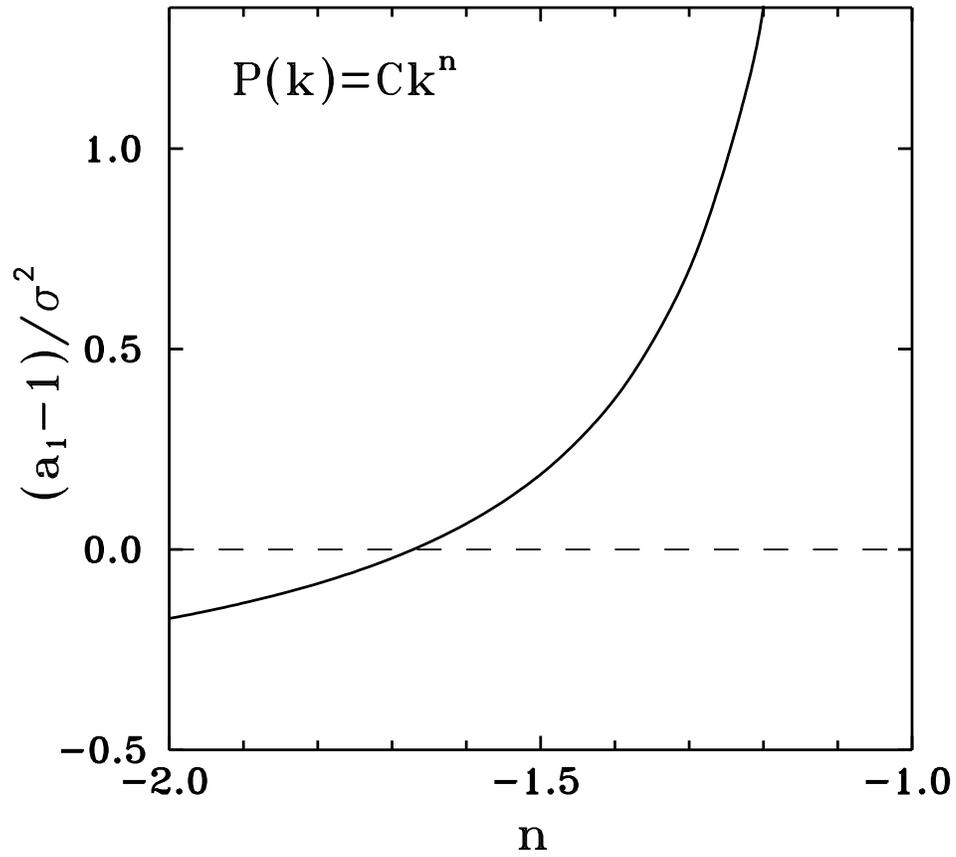}
\end{center}
    \caption{The weakly nonlinear correction to the coefficient $a_{1}$
    divided by the linear variance $\sigma^{2}$ for scale-free power
    spectra and Gaussian smoothing in the observationally most interesting
    range of the spectral index $n$.}
\label{dva1}
\end{figure}

\begin{figure}[p]
\begin{center}
    \leavevmode
    \epsfxsize=3in
    \epsfbox[155 77 455 765]{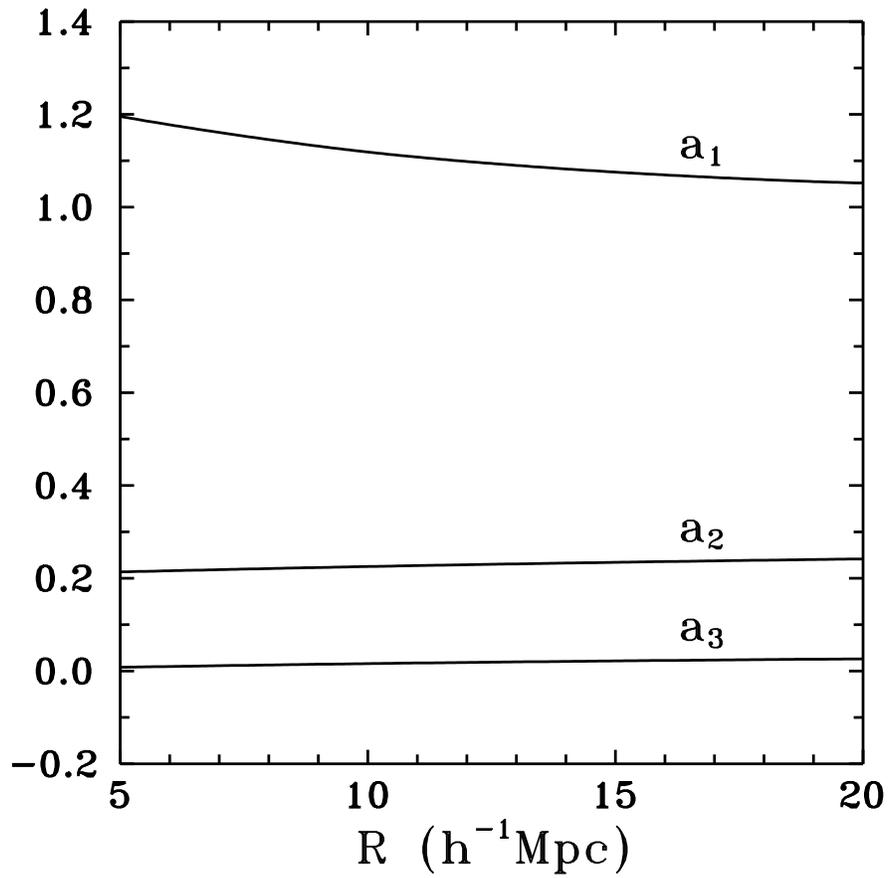}
\end{center}
    \caption{The coefficients $a_{1}$, $a_{2}$ and $a_{3}$ for the
    standard CDM spectrum normalized to (top-hat) $\sigma_{8} = 1$ in the
    weakly nonlinear range of Gaussian smoothing scales. }
\label{cdm}
\end{figure}

\end{document}